\documentclass[10pt,epsfig]{article}
\usepackage{graphicx}
\usepackage{array}
\usepackage{amsmath,amssymb}
\usepackage[latin1]{inputenc}

\begin{document}

{\Large {\bf Towards a $Z_3$-graded approach to quarks' symmetries }}
\vskip 0.4cm

{\centerline{ {\large {\bf Richard Kerner$^a$ and Jerzy Lukierski$^{b, c}$ }}}
\vskip 0.3cm

{\small { a : Laboratoire de Physique Th\'eorique de la Mati\`ere Condens\'ee - CNRS URA 7600 

 Sorbonne-Universit\'e, B.C. 121, 4 Place Jussieu 75005 Paris, France }}

{\small { b: Institute of Theoretical Physics, Wroc{\l}aw University, 
 Plac Maxa Borna 9,  Wroc{\l}aw, Poland }}

{\small { c: Bogolubov Laboratory of Theoretical Physics, Joint Inst. for Nuclear Research,

141980 Dubna (Moscow Region), Russia }}

{\it e-mail a: richard.kerner@upmc.fr}
 \; \; \; \; {\it e-mail b: jerzy.lukierski@ift.uni.wroc.pl }


\begin{abstract}

Colour $SU(3)$ group is an exact symmetry of Quantum Chromodynamics, which describes strong interactions
between quarks and gluons. Supplemented by two  internal symmetries, $SU(2)$ and $U(1)$, it serves
as the internal symmetry of the Standard Model, describing as well the  electroweak interactions 
 of quarks and leptons. The colour
 $SU(3)$ symmetry is exact, while two other symmetries are broken by means of the Higgs-Kibble mechanism.
The three colours and fractional quarks charges with values $1/3$ and $2/3$ suggest that the cyclic 
group $Z_3$ may play a crucial role in quark field dynamics.

In this paper we consequently apply the $Z_3$ symmetry to field multiplets describing colour quark fields.
Generalized Dirac equation for coloured $12$-component spinors is introduced and its properties are discussed.
Imposing $Z_3$-graded Lorentz and Poincar\'e  covariance leads to enlargement of 
  quark fields multiplets and incorporates additional 
 $Z_2 \times Z_3$ symmetry  which leads to the appearance of three generations (families) of  distinct  quark doublets.

\end{abstract}

\section{Introduction}
\vskip 0.3cm 
\indent
According to present knowledge, the ultimate undivisible and undestructible
constituents of matter, called atoms by ancient Greeks, are in fact
the elementary particles named {\it quarks}, carrying fractional electric and  baryonic charges which are 
 conserved in Standard Model under any circumstances.

What can be regarded as a mysterious feature is the fact that {\it freely propagating} quarks
have never been directly observed. However, the {\it deep inelastic scattering} experiments in which the electrons
with extremely high energies penetrate inside the nucleons and scatter on quarks, reveal the presence of almost point-like
constituents which seem to be almost free as long as they are confined inside the protons or neutrons.
When very high energy nucleons collide, they merge for a short while, and quarks 
interact via {\it gluon} exchange. The gluons are supposed to be the quanta of a {\it color} $SU(3)$ gauge field. 
Besides, quarks interact also via electroweak gauge fields.

In Quantum Chromodynamics (QCD) quarks are considered as fermions, endowed  with spin $\frac{1}{2}$.
Only {\it three quarks} or {\it three anti-quarks}
can coexist inside a fermionic baryon (respectively, anti-baryon), and a pair 
{\it quark-antiquark} can form a meson with integer spin.
Besides,  baryons are  composed of three quarks (anti-quarks) each with different colour.

 Six doublets of quarks are organized in three generations.
Two quarks in the first generation, 
$u$ and $d$ (``up" and ``down"), can be considered as two states of a more general object,
just like proton and neutron  are two isospin states of a nucleon.
 While  in baryons there is place for {\it two} quarks in the same $u$-state or $d$-state,  three $u$ or $d$ quarks in  composite
 hadron are not allowed.

The gauge gluon fields that intermediate strong interactions between quarks should carry colors.
  In fact, the totally 
``colorless" quadratic combinations of quarks  do not interact strongly with quarks, what  means that the condition 
$R {\bar{R}} + B {\bar{B}} + G {\bar{G}} = 0$, 
 where R(red), B(blue) and G(green)  denote coloured quarks,
 does  not affect the QCD strong forces.
  Eight bilinear combinations listed below describe the  octet of gluons

$$\frac{1}{\sqrt{2}} \; (R {\bar{B}} + B {\bar{R}}), 
\; \; \; \frac{1}{i \sqrt{2}} (R {\bar{B}} - B {\bar{R}}), \; \; \; \frac{1}{\sqrt{2}} \; (R {\bar{G}} + G {\bar{R}}), 
\; \; \; \frac{1}{i \sqrt{2}} (R {\bar{G}} - G {\bar{R}}),$$
$$\frac{1}{\sqrt{2}} \; (B {\bar{G}} + G {\bar{B}}), 
\; \; \; \frac{1}{i \sqrt{2}} (B {\bar{G}} - G {\bar{B}}), \; \; \; 
\frac{1}{\sqrt{2}} \; (R {\bar{R}} - B {\bar{B}}), 
\; \; \; \frac{1}{ \sqrt{6}} (R {\bar{R}} +  B {\bar{B}} - 2 G {\bar{G}}),$$
The above combinations can be linked with the choice of eight traceless Gell-Mann $3 \times 3$ matrices, spanning the $SU(3)$ Lie algebra.

In these lecture notes we shall study a way to introduce quark dynamics with incorporated $Z_3$-covariance and consider the 
 $Z_3$-graded generalization of relativistic symmetries.
As basic field equation describing coloured quarks we introduce 12-component colour Dirac equations, describing in alternative way the
quarks with three colours. The proposed $Z_3$-graded free quark equations imply that coloured quarks do not appear
asymptotically; one observes asymptotically only colourless composites, three-linear in the case of baryons and
bilinear for mesons. In colour Dirac equation the Lorentz and colour symmetries are entangled, not described by single
tensor products of  Lorentz spin and SU(3) colour matrices.

The plan of our presentation is the following. In Sect.~2 we recall the way of obtaining Dirac equation  as $Z_2$-covariant doublet
of two-component Pauli equations. In Sect.~3 we replace discrete $Z_2$ maps by
 discrete  $Z_2 \times Z_3$ transformations acting 
 on indecomposable sextet of  Pauli equations describing our basic colour Dirac equations.
 The $Z_3$-covariance modifies the dynamics of quark fields components - any component 
 of coloured Dirac equation satisfy modified
  Klein-Gordon equations, which are of the sixth order. In Sect.~4  we demonstrate how such new quarks dynamics permits to
explain algebraically the asymptotic nonobservability of coloured quarks. 
In order to incorporate fully $Z_3$-covariance of quark dynamics recently there were proposed the $Z_3$-graded
generalization of  Lorentz algebra \cite{RKJL2019} and its extension to $Z_3$-graded Poincar\ '{e} algebra
 \cite{Kerner2019B}.  In Sect.~6  we show that the formulation covariant  under standard Lorentz transformations 
 requires the consideration of doublets of colour fields called Lorentz doublets \cite{Kerner2019B};   we show also how one can introduce
the action principle for  the pairs of colour Dirac fields describing these Lorentz doublets. Last Sect.~7 contains final remarks and outlook. In particular we discuss there 
the option that coloured quark fields satisfy ternary generalization of canonical commutation relations, what leads to the
 $Z_3$-graded exclusion principle for coloured quarks (maximally only two identical colour quark states can coexist in
physical quark composites).

\section{From Pauli spinors to Dirac's equation}
\vskip 0.3cm
\indent
After the discovery of spin of the electron (the Stern-Gerlach experiment, 
Pauli understood that a Schroedinger equation involving only one complex-valued wave function
is not enough to take into account this new degree of freedom.
He proposed then to describe the dichotomic spin variable by introducing a two-component
function forming a column on which hermitian matrices can act as linear operators.

The basis of complex traceless $2 \times 2$ hermitian matrices
 contains just three elements, since then known as {\it Pauli matrices} that can be arranged in a matrix-valued $3$-vector
 ${\boldsymbol{\sigma}} = [ \sigma_1, \sigma_2, \sigma_3]:$
$$ \sigma_1 = \begin{pmatrix} 0 & 1 \cr 1 & 0 \end{pmatrix}, \; \; \; \sigma_2 = \begin{pmatrix} 0 & -i \cr i & 0 \end{pmatrix}, \; \; \; 
\sigma_3 = \begin{pmatrix} 1 & 0 \cr 0 & -1 \end{pmatrix}$$  
\noindent                                                                                                                                                                                                                                                                                                                                                                                                                                                                                   
The three Pauli matrices multiplied by $\frac{i}{2}$
span the three dimensional Lie algebra: let  $\tau_k = \frac{i}{2} \sigma_k,$ then
$$ \left[ \tau_1 , \tau_2 \right] = \tau_3,  \; \; \left[ \tau_2 , \tau_3 \right] = \tau_1,  \; \; 
\left[ \tau_3 , \tau_1 \right] = \tau_2.$$
They also form the Clifford algebra related to the Euclidean $3$-dimensional metric:
$$\sigma_i \sigma_k + \sigma_k \sigma_i = 2 \delta_{ik} \; {\mbox{l\hspace{-0.55em}1}}_2 $$
The simplest linear relation between the operators of energy, mass and momentum acting on a column
vector (called {\it a Pauli spinor}) would read then:
\begin{equation}
\begin{pmatrix} E & 0 \cr 0 & E \end{pmatrix} \; \begin{pmatrix} \psi^1 \cr \psi^2 \end{pmatrix} 
= \begin{pmatrix} mc^2 & 0 \cr 0 & mc^2 \end{pmatrix} \; \begin{pmatrix} \psi^1 \cr \psi^2 \end{pmatrix} 
+ c \; {\boldsymbol{\sigma}} \cdot {\bf p} \; \begin{pmatrix} \psi^1 \cr \psi^2 \end{pmatrix},
\label{relPauli1}
\end{equation} 
${\rm where} \; \; \; {\boldsymbol{\sigma}} \cdot {\bf p} = \sigma_1 \; p^1 + \sigma_2 \; p^2 + \sigma_3 \; p^3 =
\begin{pmatrix} p^3 & p^1 - i \, p^2 \cr p^1 + i \; p^2 & - p^3 \end{pmatrix}.$
We can write (\ref{relPauli1}) in a simplified manner, denoting the Pauli spinor by one letter
$\psi$ and treating the unit matrix symbolically like a number:
\begin{equation}
E \; \psi = m c^2 \, \psi + c \; {\boldsymbol{\sigma}} \cdot {\bf p} \; \psi.
\label{relPauli2}
\end{equation}
This  equation is not invariant under Lorentz transformations. Indeed, by iterating, i.e.
taking the square of this operator, we arrive at the following relation
between the operators of energy and momentum and the mass of the particle:
\begin{equation}
E^2 = m^2 c^4 + 2 \, m c^3 \mid {\bf p} \mid^2 {\boldsymbol{\sigma}} \cdot {\bf p} + c^2 {\bf p}^2,
\label{relPauli3}
\end{equation}
instead of the relativistic relation 
\begin{equation}
E^2 - c^2 \, {\bf p}^2 = m^2 c^4.
\label{Esquare}
\end{equation}
The double product in the expression for the energy squared can be removed if one
introduces a second Pauli spinor satisfying a smilar equation, and intertwining the two spinors. 
So let us denote the first Pauli spinor by
$\psi_{+}$ and the second one by  $\psi_{-}$, and let them satisfy the following system of coupled equations:
\begin{equation}
E \; \psi_{+} = m c^2 \, \psi_{+} + {\boldsymbol{\sigma}} \cdot {\bf p} \; \psi_{-}, \; \; \; \; 
E \; \psi_{-} = - m c^2 \, \psi_{-} + {\boldsymbol{\sigma}} \cdot {\bf p} \; \psi_{+}, 
\label{Paulidouble}
\end{equation}
(by the way, here $-1 = e^{i \pi}$, a complex number!
which coincides with the relativistic equation for the electron found by Dirac a few years later.

 The relativistic invariance is now manifest:  due to the negative mass 
term in the second equation, the iteration leads to the separation of variables, and all the
components satisfy the desired relation 
\begin{equation}
 [ E^2  - c^2 \, {\bf p}^2 ] \psi_{+} = m^2 c^4 \; \psi_{+}, \; \; \; 
[ E^2  - c^2 \, {\bf p}^2 ] \, \psi_{-} = m^2 c^4 \; \psi_{-}.
\label{Dirac1}
\end{equation}
In a more appropriate basis the Dirac equation becomes manifestly relativistic:
$ \left[ \gamma^{\mu} p_{\mu} - mc \right] \; \psi = 0,$  with $p_0 = \frac{E}{c}$, 
\begin{equation}
\gamma^0 = \sigma_3 \otimes {\mbox{l\hspace{-0.55em}1}}_2 = 
 \begin{pmatrix} {\mbox{l\hspace{-0.55em}1}}_2 & 0 \cr 0 & - {\mbox{l\hspace{-0.55em}1}}_2 \end{pmatrix},
\; \; \; \gamma^k = (i \sigma_2) \otimes \sigma^k =
\begin{pmatrix} 0 & \sigma^k \cr - \sigma^k & 0 \end{pmatrix}.
\label{gammasmall}
\end{equation}
The so defined Dirac's gamma-matrices span the Clifford algebra related to the Minkowskian metric tensor:
$$\gamma^{\mu} \gamma^{\nu} + \gamma^{\nu} \gamma^{\mu} = 2 g^{\mu \nu} \; {\mbox{l\hspace{-0.55em}1}}_4 $$
with $g^{\mu \nu} = diag [1, -1, -1, -1 ],$ while the six anti-symmetric matrices
$$\Sigma^{\mu \nu} = \frac{1}{4} \left[ \gamma^{\mu} \gamma^{\nu} - \gamma^{\nu} \gamma^{\mu} \right] $$
satisfy commutation relations characterizing the Lorentz group.

The side effect of this modification was the presence of solutions with negative mass, which
led Dirac to the conclusion that the ``holes" in the sea of such solutions could be interpreted
as electrons with the same mass as the usual ones, but with opposite charge.

 Now we have clearly the invariance under the product group $Z_2 \times Z_2$. The first $Z_2$
factor is relative to the half-integer dichotomic spin variable, the second $Z_2$ group is related
to this new universal symmetry, called charge conjugation, stipulating the existence of anti-particles.
The four-component complex function $\psi$ is a column composed of two two-component spinors,
$\xi_{\alpha}$ and $\chi_{{\dot{\beta}}}$, 
which are supposed to transform under two non-equivalent representations of the $SL(2, {\bf C})$ group: 
\begin{equation}
\xi_{\alpha'} = S^{\alpha}_{\alpha'} \xi_{\alpha}, \; \; \; \chi_{{\dot{\beta}}'} = S^{{\dot{\beta}}}_{{\dot{\beta}}'} \chi_{{\dot{\beta}}'},
\label{xichitransform}
\end{equation}
The electric charge conservation is equivalent to the vanishing of the four-divergence of $j^{\mu}$: 
\begin{equation}
\partial_{\mu} j^{\mu} = \left( \partial_{\mu} {\psi}^{\dagger} \gamma^{\mu} \right) \psi +
{\psi}^{\dagger} \left( \gamma^{\mu} \partial_{\mu} \psi \right) = 0,\end{equation}
from which we infer that this condition will be satisfied if we have
\begin{equation}
\partial_{\mu} {\psi}^{\dagger} \gamma^{\mu} = - m {\psi}^{\dagger} \; \; {\rm and} \; \; \gamma^{\mu} \partial_{\mu} \psi = m \psi,
\label{Diracfromj}
\end{equation}
which coincides with the usual Dirac equation. 

The two coupled Pauli equations can be re-written using a matrix notation:
\begin{equation}
\begin{pmatrix} E & 0 \cr 0 & E \end{pmatrix} \, \begin{pmatrix} \psi_{+} \cr \psi_{-} \end{pmatrix}
= \begin{pmatrix} mc^2 & 0 \cr 0 & - mc^2 \end{pmatrix} \, \begin{pmatrix} \psi_{+} \cr \psi_{-} \end{pmatrix}
+ \begin{pmatrix} 0 & c \, {\boldsymbol{\sigma}} {\bf p} \cr c \; {\boldsymbol{\sigma}} {\bf p} & 0 \end{pmatrix}
\; \begin{pmatrix} \psi_{+} \cr \psi_{-} \end{pmatrix},
\label{DiracPauli}
\end{equation}
where the entries in the energy operator and the mass matrix are in fact $2 \times 2$ identity
matrices, as well as the $\sigma$-matrices appearing in the last matrix, so that in reality
the above equation represents the $4 \times 4$ Dirac equation, only in a different basis.

The system of linear equations (\ref{DiracPauli}) displays two discrete $Z_2$ symmetries: the space reflection
consisting in simultaneous change of directions of spin and momentum, 
${\boldsymbol{\sigma}} \rightarrow - {\boldsymbol{\sigma}}, \; \; \; {\bf p} \rightarrow - {\bf p},$
and the particle-antiparticle symmetry realized by the transformation  
$$m \rightarrow - m, \; \psi_{+} \rightarrow \psi_{-}, \; \; \; \psi_{-} \rightarrow \psi_{+}. $$ 
Our next aim is to extend the $Z_2 \times Z_2$ symmetry by including the $Z_3$ group which
will mix not only the two spin states and particles with anti-particles, but also the three colors. 

\section{Colour Dirac equation and free quark dynamics}
\vskip 0.3cm
\indent
Recently in \cite{RKOS2014}, \cite{Kerner2017}, \cite{Kerner2018A}, \cite{Kerner2018B} a generalization of the Dirac equation for quarks
was proposed, incorporating the color degrees of freedom via extending the discrete symmetry of the system to the  $Z_3 \times Z_2 \times Z_2$
group.

 The cyclic group $Z_3$ is generated by the third root of unity, denoted by $j = e^{\frac{2 \pi i}{3}}$, with $j^2 =   e^{\frac{4 \pi i}{3}}$,
$j^3 =1$, and $1+j+j^2 = 0$.

Just as taking into account the dichotomic half-integer spin variable, the introduction of color degrees
of freedom requires additional $Z_3$ symmetry acting on a new discrete variable taking three possible (and exclusive) values, named symbolically
``red", ``blue" and ``green". The $3 \times 3$ matrices had to be introduced, all representing {\it third roots} of the $3 \times 3$ unit matrix.
Six Pauli spinors represent three colors and three anti-colors:
{\small
\begin{equation} 
\varphi_{+} = \begin{pmatrix} \varphi_{+}^1 \cr \varphi_{+}^2 \end{pmatrix}, \; \; 
\chi_{+} = \begin{pmatrix} \chi_{+}^1 \cr \chi_{+}^2 \end{pmatrix}, \; \; \; 
\psi_{+} = \begin{pmatrix} \psi_{+}^1 \cr \psi_{+}^2 \end{pmatrix}, \; \; 
\varphi_{-} = \begin{pmatrix} \varphi_{-}^1 \cr \varphi_{-}^2 \end{pmatrix}, \; \;  
\chi_{-} = \begin{pmatrix} \chi_{-}^1 \cr \chi_{-}^2 \end{pmatrix}, \; \; 
\psi_{-} = \begin{pmatrix} \psi_{-}^1 \cr \psi_{-}^2 \end{pmatrix},
\label{sixPauli}
\end{equation}}
on which Pauli sigma-matrices act in a natural way, as on two-component column spinors. 

Let us follow the logic that led from Pauli's to Dirac's equation extending it to the colors acted upon
by the $Z_3$-group. In the expression for the energy operator (i.e. the Hamiltonian), mass terms is positive when
acting on particles, and acquires negative sign acting on anti-particles, i.e. it changes sign while intertwining
particle-antiparticle components. We shall also assume that the mass term acquires the factor $j$ when we switch
from the red component $\varphi$ to the blue component $\chi$, and $j^2$ for the green component $\psi$.
The momentum operator will be non-diagonal, as in the Dirac equation, systematically intertwining not only
particles with antiparticles, but also colors with anti-colors. 

By analogy with the pair of equations (\ref{Dirac1}) in which multiplying the mass by $-1$
led to the  anti-particle appearance, now the mass term is multiplied by the generator of the $Z_3$ group, $j$, each time the colour changes.
This yields the following set of what may be called the "colour Dirac equation":


$$E \; \varphi_{+} = mc^2 \, \varphi_{+} + c \; {\boldsymbol{\sigma}} \cdot {\bf p} \, \chi_{-}$$
$$E \; \chi_{-} = - j \; mc^2 \, \chi_{-} + c \; {\boldsymbol{\sigma}} \cdot {\bf p} \, \psi_{+}$$
$$E \; \psi_{+} = j^2 \;  mc^2 \, \psi_{+} + c \; {\boldsymbol{\sigma}} \cdot {\bf p} \, \varphi_{-}$$
$$E \; \varphi_{-} = - mc^2 \, \varphi_{-} + c \; {\boldsymbol{\sigma}} \cdot {\bf p} \, \chi_{+}$$
$$E \; \chi_{+} = j \; mc^2 \, \chi_{+} + c \; {\boldsymbol{\sigma}} \cdot {\bf p} \, \psi_{-}$$
\begin{equation}
E \; \psi_{-} = -j^2 \;mc^2 \, \psi_{-} + c \; {\boldsymbol{\sigma}} \cdot {\bf p} \, \varphi_{+}
\label{systemsix}
\end{equation}

The energy operator is obviously diagonal, and its action on the spinor-valued column-vector 
can be represented as a $6 \times 6$ operator valued unit matrix.
The mass operator is diagonal, too, but its elements represent all powers of the 
sixth root of unity $q= e^{\frac{2 \pi i}{6}}$, which are  
$$q = - j^2, \; q^2 = j, \; q^3 = -1, \; q^2 = j^2, \; q^5 = - j \; {\rm and}  \;  q^6 =1.$$

The system (\ref{systemsix}) was formulated in a basis in which the ``coloured" Pauli spinors alternate
with their antiparticles; however, if we want to put forward the colour content, it is better to choose
an alternative basis in the space of spinors. From now on, we choose the
basis in which the column of six ``colored" Pauli spinors is arranged as follows:
\begin{equation}
\left( \varphi_{+}, \varphi_{-}, \chi_{+}, \chi_{-}, \psi_{+}, \chi_{-}, \psi_{-} \right)^T. 
\label{basis1}
\end{equation}
Then the mass and momentum operators take on the following form:
{\small $$M = \begin{pmatrix} m & 0 & 0 & 0 & 0 & 0 \cr 0 & - m & 0 & 0 & 0 & 0  \cr
0 & 0 & j m & 0 & 0 & 0 \cr 0 & 0 & 0 & - j m & 0 & 0 \cr 0 & 0 & 0 & 0 & j^2 m & 0 \cr
0 & 0 & 0 & 0 & 0 & -j^2 m \end{pmatrix}, \; \; \;   
P = \begin{pmatrix} 0 & 0  & 0 & {\boldsymbol{\sigma}} \cdot {\bf p} & 0 & 0   \cr 
0 & 0 &  {\boldsymbol{\sigma}} \cdot {\bf p} & 0 & 0 & 0 \cr
0 & 0 & 0 & 0 & 0 & {\boldsymbol{\sigma}} \cdot {\bf p}  \cr 
0 & 0 & 0 & 0 & {\boldsymbol{\sigma}} \cdot {\bf p} & 0 \cr 
0 & {\boldsymbol{\sigma}} \cdot {\bf p} & 0 & 0 & 0 & 0 \cr
{\boldsymbol{\sigma}} \cdot {\bf p} & 0 & 0 & 0 & 0 & 0 \end{pmatrix}$$ }

In fact, the dimension of the two matrices $M$ and $P$  displayed  above
is $12 \times 12$: all the entries in the first one are proportional to the 
$2 \times 2$ identity matrix, so that in the definition one should read 
$\begin{pmatrix} m & 0 \cr 0 & m \end{pmatrix}$ instead of $m$,
$\begin{pmatrix} j m & 0 \cr 0 & j m \end{pmatrix}$ instead of $j \; m$, etc.
The entries in the second matrix $P$ contain $2 \times 2$ Pauli's sigma-matrices,
so that $P$ is also a $12 \times 12$ matrix. The energy operator $E$ is proportional to
the $12 \times 12$ identity matrix. 

Due to the fact that only even powers of $\sigma$-matrices are proportional to \; ${\mbox{l\hspace{-0.55em}1}}_2$,
and only the powers of circulant $3 \times 3$ circulant matrix that are multiplicities of $3$ are proportional to
\;  ${\mbox{l\hspace{-0.55em}1}}_3$, the diagonalization of the system is achieved only at the sixth iteration.
The final result is extremely simple: all the components satisfy the same sixth-order equation, 
\begin{equation}
E^6 \; \varphi_{+} = m^6 c^{12} \; \varphi_{+} + c^6 \mid {\bf p} \mid^6 \; \varphi_{+}, \; \; \; \; 
E^6 \; \varphi_{-} = m^6 c^{12} \; \varphi_{-} + c^6 \mid {\bf p} \mid^6 \; \varphi_{-}.
\label{E6varphi}
\end{equation}
and similarly for all other components.

Using a more rigorous approach the three operators can be expressed 
in terms of tensor products of matrices of lower dimensions. Let us introduce two following $3 \times 3$ matrices:
{\small \begin{equation}
B = \begin{pmatrix} 1 & 0 & 0 \cr 0 & j & 0 \cr 0 & 0 & j^2 \end{pmatrix} \; \; {\rm and} \; \; 
Q_3 = \begin{pmatrix} 0 & 1 & 0 \cr 0 & 0 & 1 \cr 1 & 0 & 0 \end{pmatrix}
\label{BQmatrices}
\end{equation} }
whose products and powers generate the $U(3)$ Lie group algebra, or the $SU(3)$ algebra if we remove the unit matrix.
Then the $12 \times 12$ matrices $M$ and $P$ can be represented as the following  tensor products:
\begin{equation}
M = m \; B \otimes \sigma_3 \otimes {\mbox{l\hspace{-0.55em}1}}_2, \; \; \; \; \; \;  
P =  Q_3 \otimes \sigma_1 \otimes  ({\boldsymbol{\sigma}} \cdot {\bf p}) 
\label{MPtensor1}
\end{equation}
with as usual, $ {\mbox{l\hspace{-0.55em}1}}_2 = \begin{pmatrix} 1 & 0 \cr 0 & 1 \end{pmatrix}, \; \; \; 
\sigma_1 = \begin{pmatrix} 0 & 1 \cr 1 & 0 \end{pmatrix}, \; \; \; 
\sigma_3 = \begin{pmatrix} 1 & 0 \cr 0 & -1 \end{pmatrix}. $  

Let us rewrite the matrix operator generating the system 
(\ref{systemsix}) 
when it acts on the column vector containing twelve components of three ``colour" fields, in the basis (\ref{basis1})
$[\varphi_{+}, \varphi_{-}, \chi_{+}, \chi_{-}, \psi_{+}, \psi_{-} ]$:
$$E \; {\mbox{l\hspace{-0.55em}1}}_3 \otimes {\mbox{l\hspace{-0.55em}1}}_2 \otimes {\mbox{l\hspace{-0.55em}1}}_2 
= m c^2 \; B \otimes \sigma_3 \otimes {\mbox{l\hspace{-0.55em}1}}_2
 +  Q_3 \otimes \sigma_1 \otimes c \, {\boldsymbol{\sigma}}\cdot {\bf p}$$
with energy and momentum operators on the left hand side, and the mass operator on the right hand side:
\begin{equation}
E \; {\mbox{l\hspace{-0.55em}1}}_2 \otimes {\mbox{l\hspace{-0.55em}1}}_3 \otimes {\mbox{l\hspace{-0.55em}1}}_2 
-  Q_3 \otimes \sigma_1 \otimes c \; {\boldsymbol{\sigma}}\cdot {\bf p}
= m c^2 \; B \otimes \sigma_3 \otimes {\mbox{l\hspace{-0.55em}1}}_2
\label{EPtogether}
\end{equation}
Like with the standard Dirac equation, let us transform this equation
so that the mass operator becomes proportional the the unit matrix. To do so, we multiply the equation
(\ref{EPtogether}) from the left by the matrix $ B^{\dagger} \otimes \sigma_3  \otimes {\mbox{l\hspace{-0.55em}1}}_2.$

Now we get the following equation which enables us to interpret the energy and the momentum as the components of 
a Minkowskian four-vector $c \; p^{\mu} = [E, \; c {\bf p}] $:
\begin{equation}
E \;  B^{\dagger} \otimes \sigma_3 \otimes {\mbox{l\hspace{-0.55em}1}}_2
- Q_2 \otimes (i \sigma_2) \otimes c \, {\boldsymbol{\sigma}}\cdot {\bf p} = 
m c^2 \;  {\mbox{l\hspace{-0.55em}1}}_3 \otimes {\mbox{l\hspace{-0.55em}1}}_2 \otimes {\mbox{l\hspace{-0.55em}1}}_2,
\label{Gammafirst}
\end{equation}
where we used the fact that under matrix multiplication,
$\sigma_3 \sigma^3 = {\mbox{l\hspace{-0.55em}1}}_2$,
$B^{\dagger} B = {\mbox{l\hspace{-0.55em}1}}_3$ and $B^{\dagger} Q_3 =  Q_2$.

The sixth power of this operator gives the same result as before,
\begin{equation}
\left[ E \;  B^{\dagger} \otimes \sigma_3 \otimes {\mbox{l\hspace{-0.55em}1}}_2
- Q_2 \otimes (i \sigma_2) \otimes c \, {\boldsymbol{\sigma}}\cdot {\bf p} \right]^6 =
\left[ E^6 - c^6 {\bf p}^6 \right] \; {\mbox{l\hspace{-0.55em}1}}_{12} = m^6 c^{12} \; {\mbox{l\hspace{-0.55em}1}}_{12}
\label{sixpower1}
\end{equation}
The equation (\ref{Gammafirst}) can be written in a concise manner using the Minkowskian indices 
and the usual pseudo-scalar product of two four-vectors as follows:
\begin{equation}
\Gamma^{\mu} p_{\mu} \; \Psi = m c \; {\mbox{l\hspace{-0.55em}1}}_{12} \; \Psi, \; \; \; 
{\rm with} \; \; p^0 = \frac{E}{c}, \; \; p^k = [ \; p^x, p^y, p^z \; ].
\label{Gammasecond}
\end{equation}
with $12 \times 12$ matrices $\Gamma^{\mu}, \; \; (\mu = 0, 1, 2, 3)$ defined as follows:
\begin{equation}
\Gamma^0 = \sigma_3 \otimes B^{\dagger} \otimes {\mbox{l\hspace{-0.55em}1}}_2, \; \; \; \; \; 
\Gamma^{k} =  i \sigma_2 \otimes  Q_2 \otimes  {\sigma}^k
\label{Gammamu}
\end{equation}

In an appropriate basis, the system (\ref{systemsix}) can be represented in a Dirac-like form as follows:
\begin{equation}
\Gamma^{\mu} p_{\mu} \Psi = mc \; {\mbox{l\hspace{-0.55em}1}}_{12} \; \Psi,
\label{Terndirac0}
\end{equation} 
where $\Psi$ is the generalized $12$-component spinor made of  $6$ Pauli spinors (\ref{sixPauli}), and the generalized
$12 \times 12$ Dirac matrices $\Gamma^{\mu}$ are constructed as follows:
\begin{equation}
\Gamma^0 = B^{\dagger} \otimes \sigma_3 \otimes {\mbox{l\hspace{-0.55em}1}}_{2}, \; \; \; 
\Gamma^i = Q_2 \otimes (i \sigma_2) \otimes \sigma^i,
\label{Gammasbig}
\end{equation}
where
\begin{equation}
B^{\dagger} = \begin{pmatrix} 1 & 0 & 0 \cr 0 & j^2 & 0 \cr 0 & 0 & j \end{pmatrix}, \; \; \; \; 
B = \begin{pmatrix} 1 & 0 & 0 \cr 0 & j & 0 \cr 0 & 0 & j^2 \end{pmatrix}, \; \; \; \; 
Q_2 = \begin{pmatrix} 0 & 1 & 0 \cr 0 & 0 & j^2  \cr j & 0 & 0 \end{pmatrix}, 
\label{BQmatrix}
\end{equation}
The two traceless matrices $B$ and $Q_2$ are both cubic roots of unit $3 \times 3$ matrix. They generate
the entire Lie algebra of the $SU(3)$ group. 

The full set of six matrices $Q_A$ and $Q^{\dagger}_B, \; A, B = 1,2,3,$ together with two diagonal
traceless matrices $B$ and $B^{\dagger}$ generated by $B$ and $Q_3$ form a special basis of the $SU(3)$ algebra \cite{Kac1994}.
They can be obtained by iteration, using the following multiplication table: 
\begin{equation}
\begin{split}
& \; \; \; \; \; \; B Q_A\!\! =\!\! j^2 Q_A B\!\! =\!\! Q_{A+1}, \; \; B^{\dagger} Q_A\!\! =\!\! j Q_A B^{\dagger}\!\! =\!\! Q_{A-1},
\\
& \; \; \; \; \; \; Q^{\dagger} B\!\! =\!\! j^2 B Q^{\dagger}_A\!\! =\!\! Q^{\dagger}_{A-1}, \; \; 
Q^{\dagger}_A B^{\dagger}\!\!=\!\! B^{\dagger} Q^{\dagger}_A\!\! =\!\! Q^{\dagger}_{A+1},
\\
& \; \; \; \; \; \; \; \; \; Q_A Q_{A-1}\!\! =\!\! j Q^{\dagger}_{A+1}, \; \; Q^{\dagger}_{A-1} Q^{\dagger}_A\!\! =\!\! j^2 Q_{A+1},
\\
&Q_A Q_{A+1}^{\dagger} \! \! = \! \! B^{\dagger}, \; \; Q_A Q_{A-1}^{\dagger} \! \! = \! \! B, 
\; \; Q_A^{\dagger} Q_{A-1} \! \! = \! \! j B^{\dagger}, \; \;
Q_A^{\dagger} Q_{A+1} \! \! = \! \! j^2 B.
\end{split}
\label{RelBQ}
\end{equation}
and of course $Q_A Q_A^{\dagger} = Q_a^{\dagger} Q_A = {\mbox{l\hspace{-0.55em}1}}_{3}.$
where the indices $A, \; A+1, \; A-1$ are always taken {modulo} $3$, 
so that e.g. $3+1 \mid_{modulo \; 3} = 4 \mid_{modulo \; 3} = 1$, etc.,
and the cube of each of the eight matrices in (\ref{RelBQ}) is the unit $3 \times 3$ matrix.

In terms of their Fourier transforms, linear differential operators of any order are represented by corresponding algebraical expressions
multiplying the Fourier transform of the unknown function. The Fourier transform of the Green function is then given by the
inverse of this expression, for example, the Fourier transform of the Green function of the Klein-Gordon operator is defined as
$$ \hat{G} (k^{\mu}) = \frac{1}{k_0^2 - {\bf k}^2 - mu^2 },$$
(with $\mu = \frac{mc}{\hbar}$). The Fourier transform of Green's function for the Dirac equation is a $4 \times 4$ matrix:
$$ {\hat{D}} (k^{\mu}) =  \frac{\gamma^{\mu} k_{\mu} + m \; {\mbox{l\hspace{-0.55em}1}}_4 }{-k_0^2 + {\bf k}^2 - m^2},$$
because quite obviously one has
$$ (\gamma^{\mu} k_{\mu} + m \; {\mbox{l\hspace{-0.55em}1}}_4) ( \gamma^{\mu} k_{\mu} - m \; {\mbox{l\hspace{-0.55em}1}}_4 ) = 
 {-k_0^2 + {\bf k}^2 - m^2} \; {\mbox{l\hspace{-0.55em}1}}_4.$$
The ternary generalization of Dirac's equation being written in the most compact form as in (\ref{Terndirac0}), in terms of Fourier
transforms it becomes
\begin{equation}
\left( \Gamma^{\mu} \, k_{\mu} - m \; {\mbox{l\hspace{-0.55em}1}}_{12} \right) \; {\hat{\Psi}} (k) = 0.
\label{Dirterfour}
\end{equation}
The sixth power of the matrix $\Gamma^{\mu} k_{\mu}$ is diagonal and proportional to $m^6$, so that we have
\begin{equation}
 \left( \Gamma^{\mu} k_{\mu} \right)^6 - m^6  \; {\mbox{l\hspace{-0.55em}1}}_{12}   = 
\left( k_0^6 - \mid {\bf k} \mid^6 - m^6 \right) \;  {\mbox{l\hspace{-0.55em}1}}_{12} =  0.
\label{Gammasixem}
\end{equation}
Now we have to find the inverse of the matrix  $\left( \Gamma^{\mu} \, k_{\mu} - m \; {\mbox{l\hspace{-0.55em}1}}_{12} \right)$. 
To this effect, let us note that the sixth-order expression on the left-hand side in (\ref{Gammasixem}) can be factorized as follows:
\begin{equation}
\left( \Gamma^{\mu} k_{\mu} \right)^6 - m^6 = \left( \left(\Gamma^{\mu} k_{\mu} \right)^2 - m^2 \right) \;
\left(  \left( \Gamma^{\mu} k_{\mu} \right)^2 - j \; m^2 \right) \; \left( \left( \Gamma^{\mu} k_{\mu} \right)^2 - j^2 \; m^2 \right).
\label{Gammafact3}
\end{equation}
The first factor is in turn the product of two linear expressions, one of which is the ternary Dirac operator:
\begin{equation}
\left( \Gamma^{\mu} k_{\mu} \right)^6 - m^6 = 
\left( \Gamma^{\mu} k_{\mu} - m \right) \;\left( \Gamma^{\mu} k_{\mu} + m \right) \;
\left(  \left( \Gamma^{\mu} k_{\mu} \right)^2 - j \; m^2 \right) \; \left( \left( \Gamma^{\mu} k_{\mu} \right)^2 - j^2 \; m^2 \right).
\label{Gammafact4}
\end{equation}
Therefore the inverse of the Fourier transform of the ternary Dirac operator is given by the following matrix:
\begin{equation}
\left[ \left( \Gamma^{\mu} k_{\mu} \right) - m \right]^{-1} = 
\frac{\left( \Gamma^{\mu} k_{\mu} + m \right) \;
\left(  \left( \Gamma^{\mu} k_{\mu} \right)^2 - j \; m^2 \right) \; 
\left( \left( \Gamma^{\mu} k_{\mu} \right)^2 - j^2 \; m^2 \right)}{\left( k_0^6 - \mid {\bf k} \mid^6 - m^6 \right)}.
\label{Dirac3inverse}
\end{equation}
It takes almost no effort to prove that the numerator can be given a more symmetric form. Taking into account that
$$\left(  \left( \Gamma^{\mu} k_{\mu} \right)^2 - j \; m^2 \right) \left( \left( \Gamma^{\mu} k_{\mu} \right)^2 - j^2 \; m^2 \right) =
\left( \Gamma^{\mu} k_{\mu} \right)^4 + m^2 \; \left( \Gamma^{\mu} k_{\mu} \right)^2 + m^4,$$
we find that
$$\left( \Gamma^{\mu} k_{\mu} + m \right) 
\left(  \left( \Gamma^{\mu} k_{\mu} \right)^2 - j \; m^2 \right)  
\left( \left( \Gamma^{\mu} k_{\mu} \right)^2 - j^2 \; m^2 \right) = $$
$$ \left( \Gamma^{\mu} k_{\mu} \right)^5 + 
m \;  \left( \Gamma^{\mu} k_{\mu} \right)^4 + m^2 \;  \left( \Gamma^{\mu} k_{\mu} \right)^3 + 
m^3 \;  \left( \Gamma^{\mu} k_{\mu} \right)^2 + m^4 \;  \left( \Gamma^{\mu} k_{\mu} \right) + m^5,$$
so that the final expression can be written in a concise form as
\begin{equation}
\left[ \left( \Gamma^{\mu} k_{\mu} \right)^6 - m^6 \right]^{-1} = 
\frac{{\displaystyle \sum_{s=0}^5} m^s \; \left( \Gamma^{\mu} k_{\mu} \right)^{(5-s)}}{\left( k_0^6 - \mid {\bf k} \mid^6 - m^6 \right)}.
\label{Dirac3invsymb}
\end{equation}

In the massless case, the operator equation whose Green's function we want to evaluate, reduces to
$$ \left[ \frac{1}{c^6} \frac{\partial^6}{\partial t^6} - \left( \frac{\partial^2}{\partial x^2} + \frac{\partial^2}{\partial y^2}
+ \frac{\partial^2}{\partial z^2} \right)^3 \right] \; G (t, {\bf r}) = \delta^4 (x) = \delta(ct) \delta(x) \delta(y) \delta(z).$$
Using the Fourier transformation method, we can write:
\begin{equation} 
\left[ \frac{\omega^6}{c^6} - \mid {\bf k} \mid^6 \right] \; {\hat{G}} (k_0, {\bf k}) = 1, \; \; \; \; {\rm where} \; \; k_0 = \frac{\omega}{c},
\label{FourierGreen}
\end{equation}
from which we get 
\begin{equation}
{\hat{G}} (k_0, {\bf k} ) =  \frac{1}{k_0^6 - \mid {\bf k} \mid^6 } + \Phi (k_0, {\bf k} ),
\label{FGreen2}
\end{equation}
where $\Phi (k_0, {\bf k} )$ is a solution of the homogeneous equation, 
\begin{equation}
\left[ k_0^6 - \mid {\bf k} \mid^6 \right] \; \Phi (k_0, {\bf k} ) = 0 \; \; \; {\rightarrow} \; \; \; \Phi (k_0, {\bf k} ) = \delta (k_0^6 - \mid {\bf k} \mid^6 ).
\end{equation} 
The sixth-order polynomial $k_0^6 - \mid {\bf k} \mid^6 $ can be split into the product of three second-order factors as follows:
\begin{equation}
k_0^6 - \mid {\bf k} \mid^6 = (k_0^2 - \mid {\bf k} \mid^2) \; (k_0^2 - j \mid {\bf k} \mid^2) \; (k_0^6 - j^2  \mid {\bf k} \mid^2),
\label{kthree}
\end{equation}
each of which being a product of two linear expressions with opposite signs of $\mid {\bf k} \mid $:
$$ (k_0^2 - \mid {\bf k} \mid^2) = (k_0 + \mid {\bf k} \mid ) \,(k_0 - \mid {\bf k} \mid ), $$ 
$$ (k_0^2 -  j \, \mid {\bf k} \mid^2) = (k_0 + j^2 \; \mid {\bf k} \mid ) \,(k_0 - j^2 \; \mid {\bf k} \mid ), $$ 
$$ (k_0^2 - j^2 \,\mid {\bf k} \mid^2) = (k_0 + j \; \mid {\bf k} \mid ) \,(k_0 - j \; \mid {\bf k} \mid ), $$ 
so that the sixth-order expression appearing in (\ref{FourierGreen}) can be decomposed into a product of six linear terms.
Let us represent the inverse of this expression appearing in (\ref{FGreen2}) as a sum of three fractions with second-order
expressions in their denominators:
\begin{equation}
\frac{1}{k_0^6 - \mid {\bf k} \mid^6 } = \frac{1}{3 \mid {\bf k} \mid^4} \;  \left[ \frac{1}{k_0^2 - \mid {\bf k} \mid^2 }
 + \frac{j}{k_0^2 - j \, \mid {\bf k} \mid^2 } + \frac{j^2}{k_0^2 - j^2 \; \mid {\bf k} \mid^2 } \right],
\label{inversesix}
\end{equation}
which is to be compared with the usual Fourier inverse of the d'Alembert operator:
\begin{equation}
\frac{1}{k_0^2 - \mid {\bf k} \mid^2 } = \frac{1}{2 \mid {\bf k} \mid^2} \;  \left[ \frac{1}{k_0 - \mid {\bf k} \mid }
 - \frac{1}{k_0 +  \, \mid {\bf k} \mid } \right]
\label{inversetwo}
\end{equation}
The difference in the order of the equation leads to the difference in the algebraic structure of the polynomial representig
the equation for the Fourier transform. Its inverse displays not just two, but as much as {\it six} simple poles displayed in the following figure: 
\begin{figure}[hbt!!]
\centering 
\includegraphics[width=7cm, height=5.2cm]{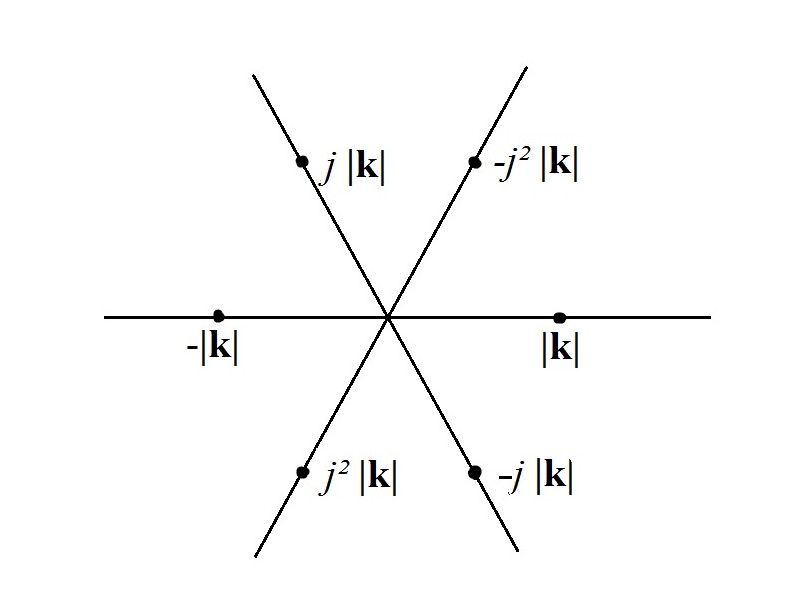}
\caption{{\small The six simple poles of the integral representation of zero-mass propagator
of the sixth-order equation}}
\label{fig:Sixpoints}
\end{figure}

Such field theories of higher order were considered by T.D.Lee and G.Wick \cite{LeeWick} and were recently an object of renewed
interest \cite{AnselmiPiva}, providing arguments showing how QFT with Lee-Wick complex poles might lead to construction of
unitary $S$-matrix.


\section{Plane wave solutions and algebraic confinement}
\vskip 0.3cm
\indent
Before discussing the properties of general solutions of our system, let us draw attention to its similarity with 
Maxwell and Dirac equations {\it in vacuo}, i.e. describing free fields with no interactions. The Maxwell equations, 
which after gauge fixing on the $4$-potential $A_{\mu}$ reduce to a system of four linear differential equations for four functions   
of $t$ and ${\bf r}$. In general, the characteristic equation of such a system should be of the fourth order; however, 
due to the particular symmetry of Maxwell's equations, the characteristic equation is of the second order (in fact, it is a fourth-order equation
which is a square of the second-order equation). This ensures the existence of a {\it single light cone} and the absence of bi-refringence in vacuum.

The same phenomenon occurs for the characteristic equation od the Dirac equation, which
is also a system of four linear differential equations for the four components of
a Dirac spinor. Like in the case of Maxwell's equations, the diagonalization of the Dirac system occurs already at the level
of the second order Klein-Gordon equation, while the characteristic equation of fourth order obtained from the determinant of the system
is the square of the second-order expression. The four independent solutions are obtained by taking two possible signs of the
square root of the determinant expression.

In our case we have as much as twelve linear equations imposed on twelve independent
functions of time and space; therefore, one could expect a twelvth-order characteristic
equation. However, due to the particular properties of Pauli's matrices and the symmetry
of our system (\ref{Terndirac0}), the diagonalization occurs already at sixth iteration:
\begin{equation}
( \Gamma^{\mu} p_{\mu} )^6 = (p_0^6 - {\bf p }^6) \;  {\mbox{l\hspace{-0.55em}1}}_{12} = m^6 c^6 \; {\mbox{l\hspace{-0.55em}1}}_{12}
\label{dispersix}
\end{equation}
This expression on the left-hand side is not manifestly relativistic invariant, but it represents a unique light cone multiplied by a positive form-factor:
\begin{equation}
(p_0^6 - {\bf p }^6) = (p_0^2 - {\bf p}^2)(j p_0^2 - {\bf p}^2)(j^2 p_0^2 - {\bf p}^2) = 
(p_0^2 - {\bf p}^2)(p_0^4 + p_0^2 {\bf p}^2 + \mid {\bf p} \mid^4).
\label{pcone}
\end{equation}
The characteristic equation of this system is obtained by taking the formal determinant of the $12 \times 12$ matrix operator;
we get
\begin{equation}
{\rm det} (\Gamma^{\mu} p_{\mu}) = (p_0^6 - \mid {\bf p} \mid^6 )^2 = m^{12} c^{12},
\label{detGamma}
\end{equation}
which is the square of the dispersion relation (\ref{pcone}).
  
In coordinate representation $\Psi = \Psi (x^{\mu})$, and the four-momentum operator $p_{\mu}$ becomes $p_{\mu} = -i \hbar \partial_{\mu}$, 
and the system of equations satisfied by the 12-dimensional wave function $\Psi (x^{\mu})$ is 
\begin{equation}
-i \hbar \; \Gamma^{\mu} \, \partial_{\mu} \, \Psi = m c \Psi
\label{TernDirac}
\end{equation}

As any linear system with constant coefficients, (\ref{systemsix}) representing twelve coupled linear equations, must have a basis
of twelve independent solutions. General solutions are then given by their linear combinations whose
coefficients depend on the choice of initial conditions.

As usually in the case of linear systems, we expect the solutions to be of general form 
\begin{equation}
f(x^{\mu}) = e^{k_{\mu} x^{\mu}} \; \; \; {\rm with} \; \; k_{\mu} \; \; {\rm satisfying} \; \; k_0^6 - \mid {\bf k} \mid^6 = - \hbar^6 m^6
\label{elemsol}
\end{equation}
Denoting $k^0 = \omega/c$, general solution can be written as $e^{\omega t - {\bf k} \cdot {\bf r}}$, ${\bf r} = [x, y, z]$:

The dispersion relation in (\ref{elemsol}) is invariant under the action of $Z_2 \times Z_3 = Z_6$ symmetry, because to any solution with given real
$\omega$ and ${\bf k}$ one can add solutions with $\omega$ replaced by  $j \omega$ or $j^2 \omega$, $j {\bf k}$ or $j^2 {\bf k}$, as well as $- \omega$; 
there is no need to introduce also $- {\bf k}$ instead of ${\bf k}$ because the vector ${\bf k}$ can take on all possible directions covering the unit sphere.

The nine complex solutions with positive frequency $\omega$ as well as with $j \; \omega$ and $j^2 \; \omega$ obtained by the action of the $Z_3$-group 
can be displayed in a compact manner in form of a $3 \times 3$ matrix. The inclusion of the essential $Z_2$-symmetry ensuring the existence of 
{\it anti-particles} leads to the nine similar solutions with negative $\omega$. This however will produce $18$ solutions, of whiich only $12$
can be linearly independent. We choose them in a symmetrical manner, displayed in two $3 \times 3$ matrices with vanishing diagonal terms:
\begin{equation}
\begin{pmatrix}  0 & e^{\omega\,t + j {\bf k \cdot r}} & e^{\omega\,t + j^2 {\bf k \cdot r}} \cr 
e^{j \omega\,t + {\bf k \cdot r }} & 0 & e^{j \omega\,t + j^2 {\bf k \cdot r}} \cr
e^{j^2 \omega\,t + {\bf k \cdot r}}& e^{j^2 \omega\,t + {j \bf k \cdot r}}  & 0 \end{pmatrix}, \; \; \; \; \; \; 
\begin{pmatrix} 0 & e^{-\omega\,t - j {\bf k \cdot r}} & e^{-\omega\,t - j^2 {\bf k \cdot r}} \cr 
e^{-j \omega\,t-{\bf k \cdot r }} & 0 & e^{-j \omega\,t - j^2 {\bf k \cdot r}} \cr
e^{-j^2 \omega\,t -{\bf k \cdot r}}& e^{-j^2 \omega\,t - j {\bf k \cdot r}}  & 0 \end{pmatrix}
\label{twomatrices}
\end{equation}

We can also produce similar matrix with six real functions, taking appropriate linear combinations as follows:
{\small
\begin{equation}
\begin{pmatrix}   0 &   e^{\omega\,t - \frac{{\bf k \cdot r}}{2}}
\, \cos ({\bf K} \cdot {\bf r}) &  e^{\omega\,t - \frac{{\bf k \cdot r}}{2}} \, \sin ({\bf K} \cdot {\bf r}) \cr 
 e^{- \frac{\omega\,t}{2}+{\bf k \cdot r }} \, \cos \Omega \, t & 0 &
e^{- \frac{\omega\,t}{2}- \frac{{\bf k \cdot r}}{2}} \, \cos (\Omega \, t  - {\bf K} \cdot {\bf r}) \cr
e^{- \frac{\omega\,t}{2} +{\bf k \cdot r}} \, \sin \Omega \, t & 
e^{- \frac{\omega\,t}{2}- \frac{ {\bf k \cdot r}}{2}} \, \sin (\Omega \, t - {\bf K} \cdot {\bf r}) & 0 
\end{pmatrix},
\label{rmatrixsix}
\end{equation} }

where $\Omega=\frac{\sqrt{3}}{2} \, \omega$ and  ${\bf K}=\frac{\sqrt{3}}{2}{\bf k}$; a similar matrix,
of course, can be produced for the alternative choice with negative $\omega$ .

All these solutions contain damping or exploding exponential factors, and cannot propagate farther than a few wavelengths.
However certain quadratic or cubic products can be chosen in such a way that these nasty factors will mutually cancel
leaving a freely propagating sinusoidal wave. Foir example, choosing two solutions from basis different than the one
displayed in (\ref{twomatrices}), namely 
\begin{equation}
e^{j \omega t + j^2 k x} \cdot e^{-j^2 \omega t - j kx} =
e^{(j-j^2) \; \omega t - (j^2 -j) \; kx} = e^{i (\Omega t - K x)},
\label{doublerun}
\end{equation} 
with $\Omega = \frac{\sqrt{3}}{2} \omega, \; \; \; \; K = \frac{\sqrt{3}}{2} k.$ 
An alternative choice of two elementary solutions leads to a free wave running in the opposite
direction:
$$e^{-j^2 \omega t - j^2 kx} \cdot e^{j \omega t + j kx} = e^{i ( \Omega t + K x)}.$$ 
This are the unique quadratic combinations yelding a running wave in both directions. This suggests that quark-anti-quark
states behave as Lorentz scalars.

We face different situation if we consider possibilities of producing a running wave with cubic expressions involving solutions
displayed in (\ref{rmatrixsix}) and their conjugates with negative $\omega$'s. Let us produce the following basis of real
solutions of (\ref{systemsix}):
$$ F_1 = e^{- \frac{\omega t}{2} + kx} \; \sin \Omega t, \; \; \; \; \; \; \;
 F_2 =e^{- \frac{\omega t}{2} + kx} \; \cos \Omega t, $$
$$G_1 = e^{\omega t - \frac{kx}{2}} \; \sin Kx, \; \; \; \; \; \; \; 
G_2 = e^{\omega t - \frac{kx}{2}} \; \cos Kx, $$
\begin{equation}
H_1 = e^{- \frac{\omega t}{2} - \frac{kx}{2}} \; \sin (\Omega t - Kx), \; \; \; \; \; 
H_2 = e^{- \frac{\omega t}{2} - \frac{kx}{2}} \; \cos (\Omega t - Kx).
\label{FGH12bis}
\end{equation} 
 Neither of the six functions above can represent a freely propagating wave: even the last two functions,
$H_1$ and $H_2$ contain, besides the running sinusoidal waves, the real exponentials which have a damping effect.
(The wave cannot penetrate distances greater than a few wavelengths, and can last only for times comparable
with few oscillations). However, we shall show that certain {\it cubic} expressions can represent
a freely propagating wave, without any damping factors. Taking a closer look at the six
solutions displayed in (\ref{FGH12bis}), we see that the only way to get rid of the real exponents 
present in all those functions, but different damping factors, is to form cubic expressions
constructed with three functions labelled with three {\it different} letters. Here is the
exhaustive list of {\it eight} admissible cubic combinations: 
$$F_1 \; G_1 \; H_1, \; \; \; \; F_1 \; G_1 \; H_2, \; \; \; \; F_1 \; G_2 \; H_1, \; \; \; \; F_1 \; G_2 \; H_2; $$
\begin{equation}
F_2 \; G_1 \; H_1, \; \; \; \; F_2 \; G_1 \; H_2, \; \; \; \; F_2 \; G_2 \; H_1, \; \; \; \; F_2 \; G_2 \; H_2. 
\label{eightFGH}
\end{equation}
But these expressions still contain, besides running waves with double frequency $2 \Omega$,
undesirable functions like $\sin \Omega t$ or $\cos K x$. To take an example, we have
$$F_1 \, G_2 \, H_2 = \sin \Omega t \; \cos K x \; \cos (\Omega t - K x) =$$
$$\frac{1}{2} \; \left[ \sin (\Omega t + Kx) + \sin (\Omega t - K x) \right] \; \cos (\Omega t - Kx)=$$
$$\frac{1}{4} \; \sin (2 \Omega t - 2 K x) + \frac{1}{4} \; \sin (2 \Omega t) 
+ \frac{1}{4} \; \sin (2 K x).$$
The explicit expressions, in terms of the trigonometric functions, of the eight independent cubic expressions
displayed above, are quite cumbersome. Here we give the final result, showing that there
are only {\it two} combinations of cubic products of solutions of the generalized ternary Dirac equation
that represent running waves, which are the following:
\begin{equation}
F_1 G_2 H_2 + F_1 G_1 H_1 - F_2 G_1 H_2 + F_2 G_2 H_1 = \sin (2 \Omega t - 2 K x),
\label{running1}
\end{equation}
\begin{equation}
F_2 G_2 H_2 + F_2 G_1 H_1 + F_1  G_1 H_2 - F_1 G_2 H_1 = \cos (2 \Omega t - 2 K x).
\label{running2}
\end{equation}
The symmetry of these expressions appears better when grouped as follows:
\begin{equation}
F_1 (G_2 H_2 + G_1 H_1) + F_2 (G_2 H_1 - G_1 H_2) = \sin (2 \Omega t - 2 K x),
\label{running1b}
\end{equation}
\begin{equation}
F_2 ( G_2 H_2 + G_1 H_1) + F_1 ( G_1 H_2 - G_2 H_1) = \cos (2 \Omega t - 2 K x).
\label{running2b}
\end{equation}
Similar two running waves are produced by forming corresponding cubic combinations
of negative frequency solutions obtained by substituting $- \omega$ instead of $\omega$. We shall get then
four independent running waves, two in one direction, two in the opposite direction. This circumstance suggests
that the three-quark combnations may behave like four-component standard Dirac spinors.

\section{$Z_3$-graded extension of $D=4$ Lorentz and Poincar\'e algebra}
\vskip 0.3cm
\indent
It is usually taken for granted that there is no other way of combining internal symmetries of interactions between
elementary particles and gauge fields  
with the Lorentz-Poincar\'e invariance interpreted as the fundamental symmetry of Minkowskian space-time, than
simple cartesian product of these groups. In the case of the Standard Model the overall symmetry is $G \times SO(1,3)$,
 where $G = SU(3)_{color} \times SU(2) \times U(1)$ and $SO(1,3)$ is the Lorentz group. The invariance with respect
to space-time translations $P_{\mu}$ extends $SO(1,3)$ to the full Poincar\'e group.

This statement results from the ``no-go theorem" by Coleman and Mandula, published in $1967$ \cite{Colemandula}.
However, it was shown since then that one can escape from the Coleman-Mandula theorem's constraints by changing the paradigm.
One of the ways of weakening one or more of its assumptions. it is possible to allow anticommuting generators as well as commuting
ones, which  leads to the possibility of supersymmetry. Supersymmetry consists in introduction 
of anticommuting symmetry generators which transform in the (12,0) and (0,12) (i.e.  spinor) representations 
of the Lorentz group. The new symmetry generators are spinors, not scalars; therefore supersymmetry is not an 
 internal  symmetry.  It  is  rather  an  extension  of  the  Poincar\'e space-time symmetries, and as such,
 has an obvious extension to space-time dimensions other than four, while the Coleman-Mandula theorem
 has no obvious extension beyond four dimensions. More details can be found in the review paper by J.D. Lykken \cite{Lykken1996}

As already observed in the previous section, the $12 \times 12$ matrices $\Gamma^{\mu}$ appearing in the coloured Dirac equation
(\ref{Dirac3}) do not satisfy the anti-commutation relations defining a 4-dimensional Clifford algebra. 
as it was the case with the $4 \times 4$ Dirac matrices $\gamma^{\mu}$. 
What can be easily seen is that their {\it third powers} are of the form \; ${\mbox{l\hspace{-0.55em}1}}_{3} \otimes \gamma^{\mu}$,
and therefore can be used to construct the usual $4$-dimensional Clifford algebra tensorized with trivial $3 \times 3$
unit matrix ${\mbox{l\hspace{-0.55em}1}}_{3}$, like in the modified version of the Standard Model proposed recently by
 I. Sogami \cite{Sogami2019}. 

Here we shall construct a spinorial representation of the $Z_3$-graded extension of the Lorentz group
following the method presented in  \cite{RKJL2019}, in which the usual Lorentz algebra ${\overset{(0)}{\cal{L}}}$ was embedded in
the $SZ_3$-graded Lorentz algebra containing three subspaces, 
\begin{equation}
{\cal{L}}_{Z_3} = {\overset{(0)}{\cal{L}}} \oplus {\overset{(1)}{\cal{L}}} \oplus {\overset{(2)}{\cal{L}}}
\label{ThreeL}
\end{equation}
The $Z_3 \otimes Z_2$ structure 
of $\Gamma^{\mu}$-matrices implies that only their sixth powers are proportional to the unit matrix ${\mbox{l\hspace{-0.55em}1}}_{12}$
(see also (\ref{detGamma})).
Thus, in order to obtain the realization of $D=4$ Lorentz algebra generators 
one can not use just two standard commutators 
\begin{equation}
 J_i = \frac{i}{2} \; \epsilon_{ijk} \left[ \Gamma^j, \Gamma^k \right], \; \; \; K_l = \frac{1}{2} \; \left[ \Gamma_l, \Gamma_0 \right] \; .
\label{JotKa}
\end{equation} 
Nevertheless, the generators $( {\overset{(0)}{J_i}}, \; {\overset{(0)}{K_l}} )$ satisfying the standard Lorentz algebra relations
(see also (\ref{modulocomm}) for $r=0, s=0$) can be defined by {triple} commutators:
\begin{equation}
\begin{split}
\\
& \left[ K_i, \left[K_j, K_k \right] \right] = \left( \delta_{ij}\delta_{kl} - \delta_{ik} \delta_{jl} \right) \; K^{(0)}_l \; .
\end{split}
\label{triplecomm}
\end{equation}
Substituting in (\ref{triplecomm}) the explicit form of $\Gamma^{\mu}$ given in (\ref{Gammasbig}), we get
\begin{equation}
\begin{split}
&J_i = - \frac{i}{2} \; Q_2^{\dagger} \otimes {\mbox{l\hspace{-0.55em}1}}_{2} \otimes \sigma_i, \; \; \; K_l = 
- \frac{1}{2} \; Q_1 \otimes \sigma_1 \otimes \sigma_l,
\\
&{\overset{(0)}{J_i}} = - \frac{i}{2} \; {\mbox{l\hspace{-0.55em}1}}_{3} \otimes {\mbox{l\hspace{-0.55em}1}}_{2} \otimes \sigma_i, \; \; \; 
{\overset{(0)}{K_l}} = - \frac{1}{2} \; {\mbox{l\hspace{-0.55em}1}}_{3}  \otimes \sigma_1 \otimes \sigma_l.
\label{NewJK}
\end{split}  
\end{equation}
\indent
In order to close the generalized Lorentz algebra (\ref{ThreeL}) where 
${\overset{(0)}{L}} \!\! =\!\! ( {\overset{(0)}{J_i}}, {\overset{(0)}{K_j}} ),$ \;   
${\overset{(1)}{L}} \!\! =\!\! ( {\overset{(1)}{J_i}}, {\overset{(1)}{K_j}} ),$ \;   
${\overset{(2)}{L}} \!\! =\!\! ( {\overset{(2)}{J_i}}, {\overset{(2)}{K_j}} ),$ one should supplement (\ref{triplecomm})
by two missing triple commutators:
\begin{equation}
\begin{split}
&\left[ J_i, \left[J_j, K_k \right] \right] = 
\left( \delta_{ij}\delta_{kl} - \delta_{ik} \delta_{jl} \right) \; K^{(2)}_l,
\\
&\left[ K_i, \left[K_j, J_k \right] \right] = \left( \delta_{ij}\delta_{kl} - \delta_{ik} \delta_{jl} \right) \; J^{(1)}_l,
\end{split}
\label{tripcommextra}
\end{equation}
where using the representation (\ref{NewJK}) we get
\begin{equation}
{\overset{(1)}{J_l}} = -\frac{i}{2} \; Q_3 \otimes {\mbox{l\hspace{-0.55em}1}}_{2} \otimes \sigma_l, \; \; \; \; 
{\overset{(2)}{K_i}} = - \frac{1}{2} \; Q_3^{\dagger} \otimes \sigma_1 \otimes \sigma_i.
\label{J1K2}
\end{equation}
The full set of $Z_3$-graded relations defining the algebra (\ref{ThreeL}) ($r, s, \; r+s$ are modulo $3$):
\begin{equation}
\begin{split}
&[ {\overset{(r)}{J_i}}, {\overset{(s)}{J_k}} ] \!\! =\!\! \epsilon_{ikl} {\overset{(r+s)}{J_l}}, \; \; 
  [ {\overset{(r)}{J_i}}, {\overset{(s)}{K_k}} ]\!\! =\!\! \epsilon_{ikl} {\overset{(r+s)}{K_l}}, 
	\\
 &[ {\overset{(r)}{K_i}}, {\overset{(s)}{K_k}} ]\!\! =\!\! - \epsilon_{ikl} {\overset{(r+s)}{J_l}}.
\end{split}
\label{modulocomm}
\end{equation} 
The remaining generators of ${\cal{L}}$ are obtained from commutators 

$[ {\overset{(1)}{K}}, {\overset{(1)}{K}} ] \simeq {\overset{(2)}{J}}$ and 
$[ {\overset{(1)}{J}}, {\overset{(1)}{J}} ] \simeq {\overset{(2)}{J}}$ :
\begin{equation}
{\overset{(2)}{J_i}} = - \frac{i}{2} \; Q_3^{\dagger} \otimes {\mbox{l\hspace{-0.55em}1}}_{2} \otimes \sigma_i, \; \; \; 
{\overset{(1)}{K_m}} = - \frac{1}{2} \; Q_3 \otimes \sigma_1 \otimes \sigma_m.
\label{J2K1}
\end{equation}
The formulae (\ref{NewJK}, \ref{J1K2}) and ({\ref{J2K1}) provide the realization of ${\cal{L}}$ which follows from
the choice (\ref{Gammasbig}) of matrices $\Gamma^{\mu}$. 

Before considering standard and generalized Lorentz covariance let us introduce the following notation:
\begin{equation}
\Gamma^{\mu}_{(A; \alpha)} = I_A \otimes \sigma_{\alpha} \otimes \sigma^{\mu}, \;  {\small A =0, 1,..,8; \; 
\alpha = 2,3; \; \mu = 0,1,2,3}\; .
\label{GammawithI}
\end{equation}
Let the $3 \times 3$ ``colour matrices" $I_A$ appearing as the first factor in (\ref{GammawithI}) be defined as follows:
$I_0 = {\mbox{l\hspace{-0.55em}1}}_{3}, \; I_r = Q_r, \; I_{r+3} = Q^{\dagger}_r, \; I_7 = B, \; I_8 = B^{\dagger}.$
Then the original $\Gamma$-matrices given by (\ref{Gammasbig}) are encoded as 
$\Gamma^0_{(8, 3)} = B^{\dagger} \otimes \sigma_3 \otimes {\mbox{l\hspace{-0.55em}1}}_{2}$
and $\Gamma^i_{(2; 2)} = Q_2 \otimes (i \sigma_2) \otimes \sigma^i$. The eight matrices with $A=1,2,...8$
with the multiplication rules given in (\ref{RelBQ}) span the ternary basis, generated by the cyclic $Z_3$-automorphism 
of the $SU(3)$ algebra (\cite{Kac1994}, Sect. $8$).

In order to get the closed formula for the action ${\cal{S}}^{(0)} \Gamma^{\mu} [{\cal{S}}^{(0)}]^{-1}$ of classical 
spinorial Lorentz symmetries generated by $L^{(0)}$, we should introduce the pairs of $\Gamma^{\mu}$-matrices 
$\Gamma^{\mu} = ( \Gamma^{i}_{(A ; 2)}, \; \Gamma^{0}_{(B;3)}) $ and $ {\tilde{\Gamma}}^{\mu} = 
(\Gamma^{i}_{(B; 2)}, \; \Gamma^{0}_{(A; 3)})$, $A \neq B$.
For any choice of $\Gamma^{\mu}$'s in (\ref{GammawithI}) we get:
\begin{equation}
[ {\overset{(0)}{J_i}}, \Gamma^{j}_{(A; \alpha)} ] = \epsilon_{ijk} \Gamma^{k}_{(A; \alpha)}, \; \; \;  
[ {\overset{(0)}{J_i}}, \Gamma^{0}_{(A; \alpha)} ] =  0,
\label{JGamma0}
\end{equation}
and the boosts ${\overset{(0)}{K_i}}$ act covariantly on doublets $\left( \Gamma^{\mu}, {\tilde{\Gamma}}^{\mu} \right)$ as follows: 
$$[ {\overset{(0)}{K_i}}, \Gamma^{j}_{(A; 2)} ] = \delta^j_i \;  \Gamma^{0}_{(A; 3)}, \;  \; \;  
[ {\overset{(0)}{K_i}}, \Gamma^{0}_{(B; 3)} ] = \Gamma^{i}_{(B; 2)},$$
\begin{equation}
[{\overset{(0)}{ K_i}}, \Gamma^{j}_{(B; 2)} ] = \delta_i^j \; \Gamma^{0}_{(B; 3)}, \; 
[{\overset{(0)}{ K_i}}, \Gamma^{0}_{(A; 3)} ] = \Gamma^{i}_{(A; 2)},
\label{Kongamma}
\end{equation}
(with $A \neq B)$, i.e. the standard Lorentz covariance requires the {\it doublet} of coloured Dirac spinors;  
In particular, the $\Gamma^{\mu}$ matrices (\ref{Gammasbig}) should be supplemented by:
\begin{equation}
{\tilde{\Gamma}}^0 = \Gamma^{0}_{(2; 3)} = Q_2 \otimes (\sigma_3) \otimes {\mbox{l\hspace{-0.55em}1}}_{2},
\; \; \; {\tilde{\Gamma}}^i = \Gamma^{k}_{(8; 2)} = B^{\dagger} \otimes i \sigma_2 \otimes \sigma^{k}.
\label{Pairgammas}
\end{equation}
One can conjecture that the pairs of $\Gamma$-matrices generated by the standard Lorentz covariance requirement 
can be used for the introduction of weak isospin doublets
of the $SU(2) \times U(1)$ electroweak symmetry. In such a way one can conclude that the internal symmetries
$SU(3) \times SU(2) \times U(1)$ of Standard Model follow from the imposition of standard Lorentz covariance
on colour Dirac multiplets.

Our next goal is to study the generalized Lorentz covariance of coloured Dirac equations, by generalization of 
standard invariance condition (\ref{Lorentz3}) 
\begin{equation}
{\tilde{\psi}} ( {\Lambda x} ) = S (\Lambda) \; \psi (x) \; S^{-1} (\Lambda )
\label{Lorentz3}
\end{equation}
and incorporating the standard $\Gamma^{\mu}$-matrices (\ref{Gammasbig}) 
into an irreducible representation of ${\cal{L}}$.
For this purpose, we should study the $18$-parameter symmetry transformation
$\Gamma^{\mu} \rightarrow {\cal{S}} \Gamma^{\mu} {\cal{S}}^{-1}$, where
\begin{equation}
{\cal{S}} = \prod_{r=0}^2 \; e^{ i \; [ \alpha^k_{(r)} {\overset{(r)}{J_k}} + \beta_{(r)}^m {\overset{(r)}{K_m}} ]},
\label{BigS}
\end{equation}
with $\alpha^k_{(0)}, \; \beta^k_{(0)}$ real, $(\alpha^k_{(1)})^{*} = \alpha^k_{(2)}, \; (\beta^k_{(1)})^{*} = \beta^k_{(2)},  \;
[ {\overset{(1)}{{J}_k}} ]^{\dagger}  = {\overset{(2)}{J_k}}$ and 
$[ {\overset{(1)}{{K}_m}} ]^{\dagger} = {\overset{(2)}{K_m}}.$
It follows that in order to obtain the closure of the faithful action of generators $(J_k^{(s)}, \; K_m^{(s)})$ ($s = 0,1,2$)
on matrices $\Gamma^{\mu}$,
one should introduce two sets $\Gamma^{\mu}_{(a)},  \Gamma^{\mu}_{\dot{a}} = (\Gamma^{\mu}_{(a)})^{\dagger}  \; (a = 1,2,...6)$ 
of coloured $12 \times 12$ Dirac matrices supplemented by Lorentz doublet partners 
$( {\tilde{\Gamma}}^{\mu}_{(a)}, {\tilde{\Gamma}}^{\mu}_{(\dot{a})}).$ 
If we choose $( {\overset{(1)}{J_k}}, {\overset{(1)}{K_m}} ) $ as given by Eqs. (\ref{J1K2}), (\ref{J2K1}), and assume that  $\Gamma^{\mu}_{(1)}$
is given by the formula (\ref{Gammasbig}), by calculating the multicommutators of 
$( {\overset{(1)}{J_i}}, {\overset{(1)}{K_l}} )$ with 
the set $\Gamma^{\mu}_{(a)}, \; (a=1,2...6)$,  we get the following sextet of $\Gamma$-matrices closed under the action of ${\overset{(1)}{L}}$ :

\begin{equation}
\begin{split}
&\Gamma^{\mu}_{(1)} = \left( \Gamma^0_{(8;3)}, \; \Gamma^i_{(2;2)} \right); \; \; 
\Gamma^{\mu}_{(4)} = \left( \Gamma^0_{(8;2)}, \; \Gamma^i_{(2;3)} \right); 
\\
&\Gamma^{\mu}_{(2)} =  \left( \Gamma^0_{(2;2)}, \; \Gamma^i_{(4;3)} \right); \; \;  
\Gamma^{\mu}_{(5)} = \left( \Gamma^0_{(2;3)}, \; \Gamma^i_{(4;2)} \right);
\\
&\Gamma^{\mu}_{(3)} =  \left( \Gamma^0_{(4;3)}, \; \Gamma^i_{(8;2)} \right); \; \; 
\Gamma^{\mu}_{(6)} = \left( \Gamma^0_{(4;2)}, \; \Gamma^i_{(8;3)} \right).
\end{split}
\label{sixGammas}
\end{equation}

The six matrices (\ref{sixGammas}) form three pairs, each of which transforms in itself under the action of the $0$-grade
subalgebra $( {\overset{(0)}{J_i}}, \; {\overset{(0)}{ K_l}} )$. The $Z_3$-graded components of ${\cal{L}}_{Z_3}$, ${\overset{(1)}{\cal{L}}}$
and  ${\overset{(2)}{\cal{L}}}$, act on these pairs transforming them into other pairs, with conjugate $Q$ and $B$ matrices. 

The realization of ${\overset{(2)}{\cal{L}}}$ sector is obtained by introducing the Hermitean-conjugate sextet
 $\Gamma^\mu_{(\dot{a})} = (\Gamma^\mu_{(a)})^{\dagger}$; further one should add 
${\tilde{\Gamma}}^\mu_{(\dot{a})} = ({\tilde{\Gamma}}^\mu_{(a)})^{\dagger}$ due to standard Lorentz covariance.
The generalized Lorentz transformations of $24$ matrices  
 $\Gamma^{\mu}_{(F)} = ( \Gamma^{\mu}_{(a)}, \Gamma^{\mu}_{(\dot{a})}; \; {\tilde{\Gamma}}^{\mu}_{(a)}, {\tilde{\Gamma}}^{\mu}_{(\dot{a})} ) $ 
will be expressed by the following generalization of the formula (\ref{Lorentz3}) 
\begin{equation}
{\cal{S}} \Gamma^{\mu}_{(F)} {\cal{S}}^{-1} = \Lambda^{\mu \; (G)}_{\; \; \nu \; (F)} \; \Gamma^{\nu}_{(G)}, 
\; \; \; \mu, \nu = 0,1,2,3; \; \; \; F,G=1,2,...,24.
\label{SLambda}
\end{equation}

 It should be recalled that the $Z_3$-grading of  Lorentz algebra was constructed explicitly  in 
 \cite{RKJL2019} only for spinorial (matrix) representations. For obtaining  full $Z_3$-graded counterpart of relativistic
symmetries described by Poincar\'{e} algebra one should introduce the orbital realization of $Z_3$-graded generalized Lorentz algebra 
  (\ref{modulocomm}), in order e.g. to apply to field-theoretic models.
Such $Z_3$-graded extension of  D=4 Poincar\'{e} algebra and its orbital realizations were constructed and discussed recently in
   \cite{Kerner2019B} by one of the present authors (RK); here we give the outline of the main ideas expressed therein.

As one can extrapolate from the realizations of standard relativistic symmetries, in construction of $Z_3$-graded orbital realizations
 the basic tool is the introduction of 12-dimensional $Z_3$-graded triplets 
(${\overset{(0)}{{p}_{\mu}}}, \; {\overset{(1)}{{p}_{\mu}}}, \; {\overset{(2)}{{p}_{\mu}}} $) 
 of  $4$-momenta and dual triplets   ($x^\mu_{(0)}, x^\mu_{(1)}, x^\mu_{(2)}$) of generalized Minkowski coordinates.
Following earlier considerations   \cite{Kerner2018B} one can introduce these triplets of  $4$-momenta  as describing three mass shells related by 
discrete $Z_3$-transformations. 
 In this way we supplement the standard real mass shell 
$p_\mu p^\mu = m^2$ ($p_\mu \equiv {\overset{(0)}{{p}_{\mu}}}$) by a pair of 
 complex-conjugated Lee-Wick mass shells   \cite{LeeWick, AnselmiPiva}
\begin{equation}
{\overset{(1)}{{p}_{\mu}}}\, {\overset{(1)}{{p}}}{}^\mu = j \; m^2, \; \; \; \; \; \; 
 {\overset{(2)}{{p}_{\mu}}}\, {\overset{(2)}{{p}}}{}^\mu = j^2 \; m^2, \; \; \; \; \; \; 
 {\overset{(1)}{{p}_{\mu}}} = ({\overset{(2)}{{p}_{\mu}}})^\star\, ,
\label{Luknext69}
\end{equation}
where
\begin{equation}
{\overset{(1)}{{p}_{\mu}}}= (p^{(1)}_0, \; {\overset{(1)}{{\bf p}}}) \; \; \; \; \; \; \; \; 
{\overset{(2)}{{p}_{\mu}}}= (p^{(2)}_0, \; {\overset{(2)}{{\bf p}}}) 
\label{Luknextb69}
\end{equation}


\begin{equation}
{\overset{(1) \; }{p_{0}}} = \pm j^2 \; \root \of{ {\overset{(1)}{{\bf{p}}^2}} + m^2}, \; \;  \; 
{\overset{(2) \; }{p_{0}}} = \pm j \; \root \of{ {\overset{(2)}{{\bf{p}}^2}} + m^2}, 
\label{Threeekzero}
\end{equation}  
 
Further one can introduce as well three versions of $4$-dimensional
 complexified space-times connected by discrete $Z_3$-transformations
\begin{equation}
z^\mu_{(s)} = ( z^0_{(s)} = j^s \, x^0_{(s)} , x^i_{(s)} ) \qquad \qquad  s=0,1,2
\label{Lukbisc}
\end{equation}
where $x^\mu_{(s)}$ are real, but  $z^\mu_{(r)}$\, ($r=1,2$)  are complex-valued and  $(z^\mu_{\ (1)})^\star = z^\mu_{\ (2)}$. 
For  s=0,1,2 one can introduce three real Minkowski scalar products in  three 8-dimensional sectors 
   (${\overset{(s)}{{p}}}_\mu , z^\mu_{(s)}$)  of
  24-dimensional generalized phase space 
	\begin{equation}
	(p,z)_{(s)} = {\overset{(s)}{{p}}_\mu} \, z^\mu_{(s)} = {\overset{(s)}{{\omega}}} \, x^0_{(s)} 	
	  - {\overset{(s)}{{p}}_i }\, x^i_{(s)}
	\label{Lukbisd}
	\end{equation}
	what allows to introduce the respective Fourier transforms.
	
	 Further, it has been shown in \cite{Kerner2019B} that one can introduce the $Z_3$-graded extension
	 	of familiar orbital realizations ${\overset{(0)}{{M}}_{\mu\nu}}$ of standard Lorentz algebra 
	\begin{equation}
	M_{\mu\nu} \equiv {\overset{(0)}{{M}}_{\mu\nu}} = x_{[ \mu}  p_{\nu ]} \ \rightarrow \ 
	({\overset{(0)}{{M}}_{\mu\nu}}, {\overset{(1)}{{M}}_{\mu\nu} }, {\overset{(2)}{{M}}_{\mu\nu}})
	\label{Lukbise}
	\end{equation}
	 where all
	 generators   ${\overset{(s)}{{M}}_{\mu\nu} }$ are constructed as bilinears of $Z_3$-graded generalized phase space coordinates
	  (${\overset{(s)}{{p}}_\mu} , z^\mu _{\, (s)}$).
		  The formula for total (orbital + spin)  $Z_3$-graded  relativistic angular  momentum 
			  ${\overset{(r)}{{\mathcal{M}}}_{\mu\nu}} = ({\overset{(r) }{{\cal{J}}_i}} , {\overset{(r)}{{\mathcal{K}}_l }}$)
			 is given by the formula
			\begin{equation}
		   {\overset{(r) }{{\cal{J}}_i}}= {\overset{(r)}{{J}}_i{}^{\!\!orb}} + {\overset{(r)}{{J}}_ i}
			\qquad
				{\overset{(r)}{{\mathcal{K}}_i }}= {\overset{(r)}{{K}}_i{}^{\!\!orb}} + {\overset{(r)}{{K}}_ i}
			\label{Lukbisf}
			\end{equation}
			where the  matrix-like ``spin'' part of generators  (${\overset{(r)}{J}_i, {\overset{(r)}{K}_i}} $) 
			were studied in  \cite{RKJL2019}.  
			 The orbital and spin parts of  the realizations  (\ref{Lukbisf}) commute with each other and
			  (${\overset{(r)}{\mathcal{J}_i}}, {\overset{(r)}{\mathcal{K}_i}} $)
			satisfy the  $Z_3$-graded  generalized Lorentz algebra relations (see (\ref{modulocomm})).
			
			Further one can extend the $Z_3$-grade relations  (\ref{modulocomm}) to $Z_3$-graded Poincar\'{e} algebra, 
			 firstly presented in  \cite{Kerner2019B}, 
			  with 18  generators  ${\overset{(s)}{{M}}_{\mu\nu}}$ and 12 generators ${\overset{(s)}{{P}}_ \mu}$:

The $Z_3$-graded extension of full Poincar\'e algebra is quite obvious \cite{Kerner2019B}. Whatever the representation we choose (spinorial
or ``orbital"), the commutation relations for the $Z_3$-graded Lorentz subalgebra remain the same. Let us denote them in case they
are part of the extended $Z_3$-graded Poincar\'e algebra by ${\overset{(r) }{{\cal{K}}_i}}$ (generalized Lorentz boosts) and ${\overset{(r) }{{\cal{J}}_i}}$
(the generalized spatial rotations), where the superscript $r = 0, 1, 2$ refers to the $Z_3$-grade of one of the three components of
$Z_3$-graded extended Lorentz algebra, and $i, k = 1,2,3$ are the $3$-space indices.

The commutation rules of the Lorentz algebra:\
\begin{equation}
[ \; {\overset{(r) }{{\cal{K}}_i}}, \; {\overset{(s) }{{\cal{K}}_k}} \; ]  = - \epsilon_{ikl} {\overset{(r+s) }{{\cal{J}}_l}}, \; \; \; \; \; 
[ \; {\overset{(r) }{{\cal{J}}_i}}, \; {\overset{(s) }{{\cal{K}}_k}} \; ]  = \epsilon_{ikl} {\overset{(r+s) }{{\cal{K}}_l}}, \; \; \; \; \;%
[ \; {\overset{(r) }{{\cal{J}}_i}}, \; {\overset{(s) }{{\cal{J}}_k}} \; ] = \epsilon_{ikl} {\overset{(r+s) }{{\cal{J}}_l}}.
\label{modulocomm1}
\end{equation}
must be now supplemented by set of commutation rules between Lorentz generators and the generators of $4$-translations, which
should also form a  $Z_3$-graded extension of usual $4$-dimensional Minkowskian translations $P_{\mu}$. Denoting them by

\begin{equation}
[ \; {\overset{(0) }{{\cal{P}}_{\mu}}}, \; \; \;  {\overset{(1) }{{\cal{P}}_{\mu}}}  \; \; \; {\overset{(2) }{{\cal{P}}_{\mu}}} ],
\label{PPcomm}
\end{equation}
with $r = 0, 1, 2$ and $\mu, \nu = 0,1 2,3$,
we impose the following $Z_3$-graded extra commutation relations:

\begin{equation}
[ \; {\overset{(r) }{{\cal{P}}_0}}, \; {\overset{(s) }{{\cal{P}}_k}} \; ]  = 0; \; \; \; \; \;
[ \; {\overset{(r) }{{\cal{P}}_i}}, \; {\overset{(s) }{{\cal{P}}_j}} \; ]  = 0,
\label{PPcomm2}
\end{equation}
\begin{equation}
[ \; {\overset{(r) }{{\cal{J}}_k}}, \; {\overset{(s) }{{\cal{P}}_0}} \; ]  = 0; \; \; \; \; \; 
[ \; {\overset{(r) }{{\cal{J}}_i}}, \; {\overset{(s) }{{\cal{P}}_k}} \; ] = \epsilon_{ikl}\; {\overset{(r+s) }{{\cal{P}}_l}}, 
\label{JPgradcomm}
\end{equation}
\begin{equation}
[ \; {\overset{(r) }{{\cal{K}}_i}}, \; {\overset{(s) }{{\cal{P}}_0}} \; ]  =  {\overset{(r+s) }{{\cal{P}}_i}}, \; \; \; \; \; 
[ \; {\overset{(r) }{{\cal{K}}_i}}, \; {\overset{(s) }{{\cal{P}}_k}} \; ]  = - \delta_{ik}  {\overset{(r+s) }{{\cal{P}}_0}}.
\label{KJPextended}
\end{equation}


In all the above relations the grades $r, s = 0,1,2$ add up modulo $3$.
Irreducible representations of the Poincr\'e algebra (and also the group, by exponentiation) are characterized by eigenvalues of its
Casimir operators, the most important of which is the mass operator $M^2 = P_{\mu} P^{^mu}$.
It turns out \cite{Kerner2019B} that in order to generalize the Casimir operator given by the square of four-momentum
we must take into account similar contributions from all possible combination of $Z_3$ grades:
\begin{equation}
{\cal{P}}^2 =  {\overset{(0)}{{\cal{P}}_{\mu}}} {\overset{(0)}{{\cal{P}}^{\mu}}} + {\overset{(1)}{{\cal{P}}_{\mu}}} {\overset{(1)}{{\cal{P}}^{\mu}}}
+ {\overset{(2)}{{\cal{P}}_{\mu}}} {\overset{(2)}{{\cal{P}}^{\mu}}} + {\overset{(0)}{{\cal{P}}_{\mu}}} {\overset{(1)}{{\cal{P}}^{\mu}}}
+ {\overset{(1)}{{\cal{P}}_{\mu}}} {\overset{(2)}{{\cal{P}}^{\mu}}} + {\overset{(2)}{{\cal{P}}_{\mu}}} {\overset{(0)}{{\cal{P}}^{\mu}}},
\label{Z3PCasimir} 
\end{equation}
It can be checked that this operator commutes with the full set of generators of the Lorentz-Poincar\'e algebra by virtue of
(\ref{PPcomm2}, \ref{JPgradcomm} and \ref{KJPextended}).

The analysis of eigenvalues of (\ref{Z3PCasimir}) as well as those of the second Casimir operator of fourth order, corresponding
to orbital spin constructed with Pauli-Lubanski operator, will be the subject of our forthcoming publications.


 At present we study the possibilities of incorporating the $Z_3$-graded Poincar\'{e} symmetries 
 (\ref{PPcomm2}-\ref{KJPextended})  as a guiding line for the introduction of $Z_3$-graded generalization of field-theoretic QCD  model,  which  incorporates complete
 quark sector of the Standard Model.

\section{The Lorentz-covariant Lagrangean for quark doublets}
\vskip 0.3cm
\indent
In this Section we shall show how the free Lagrangean for the pair of colour Dirac spinors can be made
 covariant under the standard Lorentz transformations.
Let us start by recalling the properties of color-spinor representation of the Lorentz group
 transforming   the generalized colour Dirac matrices $\Gamma^{\mu}$. According to the notation 
 proposed  in  \cite{RKJL2019},  we have:

\begin{equation}
\Gamma^{\mu} =: \left( \Gamma^0, \; \Gamma^k \right) =
\left( B^{\dagger} \otimes \sigma_3 \otimes {\mbox{l\hspace{-0.55em}1}}_2,  \; \; Q_2 \otimes (i \sigma_2)  \otimes \sigma^k \right),
\label{Gammazero}
\end{equation}
while the generators of the colour Lorentz group are given by the following six matrices, three space rotations and three
independent Lorentz boosts:
\begin{equation}
{\overset{(0)}{J_i}} = - \frac{i}{2} \; {\mbox{l\hspace{-0.55em}1}}_{3} \otimes {\mbox{l\hspace{-0.55em}1}}_{2} \otimes \sigma_i, \; \; \; \; \; \; 
{\overset{(0)}{K_l}} = - \frac{1}{2} \; {\mbox{l\hspace{-0.55em}1}}_{3}  \otimes \sigma_1 \otimes \sigma_l.
\label{NewJK2} 
\end{equation}
Taking the commutators of these generators (\ref{NewJK}) we discover that the matrices $\Gamma^{\mu}$ are transformed into another set
which we shall denote by ${\Tilde{\Gamma}}^{\mu}$:
$$ [ {\overset{(0)}{J}}_i , \; \Gamma^0 ] = 0, \; \; \; [ \; {\overset{(0)}{J}}_i  , \; \Gamma^k \;] = \epsilon_{ikm} \Gamma^m; $$
\begin{equation} 
 [ {\overset{(0)}{K}}_i , \; \Gamma^0 ] = {\tilde{\Gamma}}^i, \; \; \; [ \; {\overset{(0)}{K}}_i , \; \Gamma^k \; ] = \delta^k_i {\tilde{\Gamma}}^0;
\label{LorGamma}
\end{equation}
where ${\Tilde{\Gamma}}^{\mu}$ are given by the following tensor products:
\begin{equation}
{\tilde{\Gamma}}^{\mu} =: \left( {\tilde{\Gamma}}^0, \; {\tilde{\Gamma}}^k \right) =
\left( Q_2 \otimes \sigma_3 \otimes \; {\mbox{l\hspace{-0.55em}1}}_2,  \; \; B^{\dagger} \otimes (i \sigma_2)  \otimes \sigma^k \right),
\label{Gammatildezero}
\end{equation}
Quite obviously, when acting with the same generators on ${\tilde{\Gamma}}^{\mu}$, we come back to the initial set $\Gamma^{\mu}$,
with exactly the same relations as (\ref{LorGamma}). 

It is also easy to check that the operator ${\tilde{\Gamma}}^{\mu} \; p_{\mu}$ satisfies the same fundamental relations as the
original operator $\Gamma^{\mu} \; p_{\mu}$, namely
\begin{equation}
\left( {\tilde{\Gamma}}^{\mu} p_{\mu} \right)^6 = (p_0^6 - \mid {\bf p} \mid^6) \; {\mbox{l\hspace{-0.55em}1}}_{12}, \; \; \; \; 
{\rm det} \left( {\tilde{\Gamma}}^{\mu} p_{\mu} \right) = (p_0^6 - \mid {\bf p} \mid^6)^2.
\label{sixdet}
\end{equation}

The matrix-valued four-vector $\Gamma^{\mu}$ enter into the definition of the generalized (``colour") Dirac equation:
\begin{equation}
\Gamma^{\mu} \partial_{\mu} \; \Psi = m \Psi.
\label{Dirac3}
\end{equation}
Now we clearly see that this equation cannot be covariant under the Lorentz transformations (\ref{LorGamma}), because it does not contain
the other member of the doublet, ${\tilde{\Gamma}}^{\mu}$. To restore the symmetry, we must introduce another coloured spinor, to
be associated with the ${\tilde{\Gamma}}^{\mu}$ which form a Lorentz doublet of coloured spinors . 
 In order to make the  Lorentz doublet equations covariant under classical Lorentz transformations we should 
 introduce  mass terms
as  introducing the mixing between $\Psi$ and ${\tilde{\Psi}}$:
\begin{equation}
\Gamma^{\mu} \partial_{\mu} \; \Psi = m {\tilde{\Psi}}, \; \; \; \; {\tilde{\Gamma}}^{\mu} \partial_{\mu} \; {\tilde{\Psi}} = m \Psi,
\label{TwoDiracs}
\end{equation}
The covariance is ensured because one can show that under the same Lorentz action $\Psi$ becomes ${\tilde{\Psi}}'$ and ${\tilde{\Psi}}$ becomes $\Psi'$.

In order to produce the Lagrangian action leading to the pair of coupled colour Dirac equations (\ref{TwoDiracs}) we shall follow as
 faithfully as possible the classical Dirac scheme. The Dirac lagrangean was defined as follows:
\begin{equation}
{\cal{L}}_{Dirac} = \frac{i}{2} \left[ \left( \partial_{\mu} {\bar{\psi}} \right)  \gamma^{\mu} \psi - 
{\bar{\psi}} \gamma^{\mu} \partial_{\mu} \psi \right] + m {\bar{\psi}} \psi.
\label{ClassDirac}
\end{equation}
with Dirac conjugation defined as follows (in order to guarantee the positivity of the energy):
\begin{equation}
{\bar{\psi}} = \psi^{\dagger} \gamma^5, \; \; \; \; \; \gamma^5 = \sigma_1 \otimes \;  {\mbox{l\hspace{-0.55em}1}}_2
\label{Dirconj}
\end{equation}
The Lorentz-covariant generalization of the Lagrangean (\ref{ClassDirac}) looks as follows
\begin{equation}
{\cal{L}}_0 = \frac{i}{2} \left[ \partial_{\mu} {\bar{\tilde{\Psi}}} \Gamma^{\mu} \Psi - {\bar{\tilde{\Psi}}} \Gamma^{\mu}\partial_{\mu} \Psi
+ \partial_{\mu} {\bar{\Psi}} {\tilde{\Gamma}}^{\mu} {\tilde{\Psi}} -  {\bar{\Psi}} {\tilde{\Gamma}}^{\mu} \partial_{\mu} {\tilde{\Psi}} \right]
+ m {\bar{\tilde{\Psi}}} {\tilde{\Psi}} + m {\bar{\Psi}} \Psi.
\label{Lagranfull}
\end{equation}
One can easily check that variation with respect to $\Psi$ or ${\tilde{\Psi}}$ lead respectively to the equations (\ref{TwoDiracs}) and 
provides the conjugate equations if the variation is made with respect to ${\bar{\Psi}}$ and ${\bar{\tilde{\Psi}}}$.

What remains to be defined is the generalization of Dirac's conjugation ${\bar{\Psi}} = \Psi^+ \, \gamma_0$. 
 In order to ensure the positivity of the energy, these
conjugations should be defined as follows:
\begin{equation}
{\bar{\Psi}} = {\Psi^{\dagger}} ( B \otimes \sigma_3 \otimes \; {\mbox{l\hspace{-0.55em}1}}_2,) \; \; \; \; \;
{\bar{\tilde{\Psi}}} = {\tilde{\Psi}}^{\dagger} ( Q_2^{\dagger} \otimes \sigma_3 \otimes \; {\mbox{l\hspace{-0.55em}1}}_2).
\label{NewDiraconj}
\end{equation}
This choice of generalized Dirac conjugation for coloured spinors yields the desired result, namely:
\begin{equation}
{\bar{\Psi}} \Gamma^0 \Psi =  {\Psi^{\dagger}} ( B \otimes \sigma_3 \otimes {\mbox{l\hspace{-0.55em}1}}_2)
(B^{\dagger} \otimes \sigma_3 \otimes \; {\mbox{l\hspace{-0.55em}1}}_2) \Psi = \Psi^{\dagger} \Psi,
\label{Psiplus}
\end{equation}
and
\begin{equation}
{\bar{\tilde{\Psi}}} {\tilde{\Gamma}}^0 {\tilde{\Psi}} =  
{{\tilde{\Psi}}^{\dagger}} ( Q_2^{\dagger} \otimes \sigma_3 \otimes {\mbox{l\hspace{-0.55em}1}}_2)
(Q_2 \otimes \sigma_3 \otimes \; {\mbox{l\hspace{-0.55em}1}}_2) {\tilde{\Psi}} = {\tilde{\Psi}}^{\dagger} {\tilde{\Psi}}.
\label{Psitildeplus}
\end{equation}  

Similar construction can  be performed with two remaining doublets of $\Gamma$-matrices introduced in (\ref{sixGammas}), and 
two analogous Lagrangeans involving the pairs $(\Gamma^{\mu}_{(2)}, \; \Gamma^{\mu}_{(5)})$ and $(\Gamma^{\mu}_{(3)}, \; \Gamma^{\mu}_{(6)})$
should be added  to the action given by (\ref{Lagranfull}).

\section{Final remarks and outlook}
\vskip 0.3cm
\indent
The colour generalization of Dirac's equation proposed here with mixing of colour  and spin degrees of freedom and 
incorporating particle-antiparticle duality displays natural $Z_3 \times Z_2 \times Z_2$ symmetry. Imposing the 
$Z_3$-graded Lorentz and Poincar\'e
covariance required the enlagement of the usual $10$-dimensional algebra to its $30$-dimensional $Z_3$-graded extension,
acting on two conjugate sextuplets of generalized $12$-dimensional coloured spinors.

Although the characteristic equation of the colour Dirac operator is of sixth order, this operator will display the
same ultraviolet behavior of perturbative Feynman integrals as the usual Dirac equation, because the inverse operator's Fourier transform
shows that it behaves as $k^{-1}$ in high energy limit. The complex-valued poles lead to
 the exponentially damped solutions which cannot
propagate at distances larger than a few characteristic wavelenghts and other ones 
 exponentially blowing up.
  However, as we did show in Sect.~4  the products of three  suitable solutions  of colour Dirac equations 
  can produce freely propagating oscillating  waves.

 Such a phenomenon seems to be analogous to the properties of electrons' with opposite wave vectors
${\bf k}$ and $- {\bf k}$ couplings in Cooper pairs behaving like bosons and are the source of supraconductivity phenomenon.
By analogy, the $Z_3$-symmetry creates the situation in which the damped quark states obeying the coloured Dirac equation 
can form composite states when in momentum space
their wave vectors are ${\bf k}, \; j {\bf k}$ and $j^2 {\bf k}$ forming a propagating fermion, or coupled in pairs
with wave vectors $j {\bf k}$ and $- j^2 {\bf k}$ forming a propagating boson.

Another remarkable property of $Z_3$ symmetry to be exploited in quark model consists in a generalization of Pauli's 
exclusion principle \cite{Pauli1926}, \cite{Kerner2017}. Although quarks behave like Pauli spinors, obeying the Fermi-Dirac
statistics, they possess, beside the two-valued spin and three-valued colour also another two-valued characteristics, labeled
 $u$ and $d$ in the first family, $c$ and $s$ in the second, and $t$ and $b$ in the third one. But this new attribute is not 
exclusive like the half-integer spin or colour: it can accomodate in composite state two particles labeled identically, but not three.
Such a behavior can be modelled by a $Z_3$-graded exclusion principle making use of $j$-skew symmetry.
 In such a case  the quark wave functions' algebra
  should take values in an associative $Z_3$-graded algebra whose generators obey the following {\it ternary commutation rules};
\begin{equation}
\theta^A \theta^B \theta^C = j \; \theta^B \theta^C  \theta^A = j^2 \;\theta^C \theta^A \theta^B, \; \; \; A, B,.. = 1, 2, ...N
\label{thetas}
\end{equation}
 along with their hermitian conjugated ternary rules for conjugate generators ${\bar{\theta}}^{\dot{B}}$.

The generators $\theta^A$ are of $Z_3$ grade $1$, their hermitian conjugates ${\bar{\theta}}^{\dot{B}}$  of $Z_3$ grade $2$,
and to physical states we assign the products of generators whose $Z_3$ grade is $0$.
Briefly, the squares of generators are not zero, but their cubes vanish identically, $(\theta^A)^3 = 0$.

In such a framework, in the case of two generators, $A, B,.. = 1, 2, \; {\dot{A}}, {\dot{B}}, .. = {\dot{1}}, {\dot{2}}$,
we can choose as the only non-vanishing independent cubic combinations of $Z_3$ grade $0$ the following ones:
\begin{equation}
\theta^1 \theta^1 \theta^2, \; \; \; \theta^1 \theta^2 \theta^2, \; \; \;   
{\bar{\theta}}^{\dot{1}} {\bar{\theta}}^{\dot{1}} {\bar{\theta}}^{\dot{2}}, \; \; \; 
{\bar{\theta}}^{\dot{1}} {\bar{\theta}}^{\dot{2}} {\bar{\theta}}^{\dot{2}},
\label{thetacubes}
\end{equation}
which can describe stable fermionic quark composites $uud, \; udd, \; {\bar{u}} {\bar{u}} {\bar{d}}, \; {\bar{u}} {\bar{d}} {\bar{d}};$. 
Analogously, the independent quadratic monomials of $Z_3$ grade zero
\begin{equation}
\theta^1 {\bar{\theta}}^{\dot{2}}, \; \; \theta^2 {\bar{\theta}}^{\dot{1}}, \; \; \; 
\theta^1 {\bar{\theta}}^{\dot{1}} - \theta^2 {\bar{\theta}}^{\dot{2}},
\label{thetaquad}
\end{equation}
  correspond to bosonic quark composites $u {\bar{d}}, \; \; {\bar{u}} d, \; \; u {\bar{u}} - d {\bar{d}}.$ 

Our next aim will be to investigate the quark fields endowed with the $Z_3$-graded exclusion principle 
  and  $Z_3$-graded extended relativistic symmetries 
as  describing 
 the
dynamics of suitably modified Standard Model, in particular  modified colour sector of strong interactions 
(chromodynamics). This stage of development of our framework, with the discussion of its applicability 
 to particle physics is 
  now under consideration.  

\vskip 0.3cm
\indent
{\bf Acknowledgement}
\vskip 0.2cm
\indent
\hskip 0.3cm
One of the authors (JL) would like to express his gratitude to  Bogolubov Institute of Theoretical Physics 
  (JINR,  Dubna) for warm hospitality extended to him during his stay there in Summer of 2019.

\end{document}